\documentclass[floats,aps,showpacs]{revtex4} 
\usepackage{dcolumn}

\usepackage{epsfig}
\newcommand {\la} {\langle}
\newcommand {\ra} {\rangle}
\newcommand {\beq} {\begin{eqnarray}}
\newcommand {\eeqn} [1] {\label{#1} \end{eqnarray}}%
\newcommand {\eol} {\nonumber \\}
\newcommand {\ve} [1] {\mbox{\boldmath $#1$}}

\begin{document}
%
%

\title{Long-range behaviour in hyperspherical formalism}

\author{
N.\ K.\ Timofeyuk
}        

\affiliation{
Physics Department, University of Surrey, Guildford,
Surrey GU2 7XH, England, UK
}

\date{Received: \today}

\begin{abstract}

It is shown that   hyperspherical harmonics can be represented
in a form typical of traditional microscopic cluster models.
This allows those hyperharmonics that
are responsible for  long-range behaviour of valence
nucleons in loosely-bound nuclei to be selected. The hyperspherical cluster
model based on such hyperharmonics is tested for  
  $^5$He using a simplified description of the $^4$He core and
  two versions of the Volkov potential 
  which give $^5$He as either bound or unbound.
The study
has confirmed that it is possible to get a converged solution for
binding energy, r.m.s. radius, overlap integral and asymptotic
normalization coefficient with small
number of basis functions for bound $^5$He. 
The values obtained of these quantities are very close to those calculated
in a microscopic cluster model.

\end{abstract}

\maketitle

\section{INTRODUCTION}

Weak binding in nuclei with a large excess of neutrons or protons leads
to strong clustering and to a slow decrease of the wave function
of valence nucleons at large distances. 
To describe the processes in which such nuclei are involve
and to deduce  structure characteristics  through the study of these
processes,   long-range behaviour of valence nucleons
should be properly taken into account. Current reaction theories often 
use simplified phenomenological
nucleon-core models to include long-range behaviour relevant to reaction
mechanisms. However, a deeper understanding of nuclear structure and  
its manifestation in nuclear reactions requires  
nuclei to be considered as   many-body objects.

It is not easy nor always possible to obtain correct long-range
behaviour  from first principles. It 
can be obtained naturally  only for  
  three-body and four-body systems  using Faddeev and Faddeev-Yakubovsky
equations. For three-nucleon systems,  correct long-range behaviour
has also been achieved using pair correlated hyperspherical harmonics
\cite{Kie97},
 which  has allowed  asymptotic normalization constants
for the $n-d$ and $p-d$ systems to be calculated. 
The hyperspherical formalism has also been
applied to calculations of the overlap integral between 
triton and $^4$He \cite{Viv05}.
Despite the very large basis space used in these
calculations, the proper asymptotic behaviour of the  overlap 
$\la t\otimes p |^4$He$\ra$, given by the Whittaker function
with relevant proton separation energy,
has not been achieved at $r > 6$ fm. To restore this behaviour,
the authors solved the inhomogeneous equation for overlap integrals fro Ref.
 \cite{Tim98}. For heavier nuclei, only a few ab-initio calculations of overlap
integrals are available. Large-scale no-core
shell model calculations of the
  overlap $\la ^7$Be$\otimes p |^8$B$\ra$ \cite{Nav06} 
  have not provided the required decrease at $r > 5 $ fm so that the authors
had to match 
this overlap to the Whittaker function, or approximate it
by solutions 
of a Woods-Saxon potential  in order to use this  overlap in reaction
calculations. Three other ab-initio overlap
calculations are available within the variational Monte-Carlo  method, for  
$\la^4$He$\otimes d| ^6$Li$\ra$ \cite{Nol01a},
$\la^4$He$\otimes t| ^7$Li$\ra$ and $\la^4$He$\otimes ^3$He$| ^7$Be$\ra$
\cite{Nol01b}.
The radial behaviour of the
long-range part of these overlaps 
is imposed there prior to calculations so that
it is not determined by the differences in 
calculated energies of the nucleus and the clusters that compose it.

Many-body approaches that use Slater determinants made
of single-particle wave functions
have also to adjust the potential wells to fit the experimental energies for
valence nucleons if such wave functions are to be used in reaction
calculations. However, this procedure does not necessarily provide the
separation energy that is the difference between the total energies. Also,
it  may be inconsistent with nucleon-nucleon  (NN) interactions employed
in these approaches.

Proper long-range behaviour can be guaranteed if the total wave function
$\Psi$ is represented by an antisymmetried product of the core 
$\phi_c$ and valence $\phi_{val}$ wave functions
\beq
\Psi = {\cal A} (\phi_c \otimes \phi_{val})
\eeqn{eq1}
and then the microscopic R-matrix approach is used 
determine the function $ \phi_{val}$ \cite{Bae77}. 
At present a microscopic cluster model (MCM) of this type
is  used to predict
cross sections of astrophysically relevant low-energy reactions
\cite{pdesc}.
However, this model is only able to deal    with
 oscillator shell model core wave functions $\phi_c$ 
and can not be  used with realistic NN interactions
that reproduce  NN data.
Although in this model the separation energies are always equal to the
differences of the total binding energies,  they  often 
differ  significantly from experimental values so that one
parameter of the effective NN potential 
should be tuned. 

The problem of the long-range behaviour can be resolved within the
hyperspherical interpolation approach proposed in \cite{BZh}.
 This approach
has been designed to  describe  many-nucleon  systems near thresholds
when identified channels having a binary cluster structure are believed
to play  an important role, especially in  the nuclear
surface and beyond.
In this approach the wave function of a many-nucleon system $\Psi$
is represented by  two terms $\Psi = \Psi_1 + \Psi_2$,
 one of which 
explicitly  contains the binary channel wave function. Both
 $\Psi_1$ and $\Psi_2$   are expanded in hyperspherical basis and
a system
of equations that couples them derived by substituting the
total wave function $\Psi$
into the general  expression for the variational
principle.  
The interpolation approach does not involve matching at some surface. The 
coupling between the two components of the wavefunction is governed by the NN 
interaction through the Schr\"odinger equation. 
Earlier applications of the interpolation method included the calculation of  
 $^4$He+n elastic scattering \cite{EfrZh}. Later, 
this approach was used to study the long-range
behaviour in $^8$B and $^6$Li nuclei within a three-body model
\cite{G1}.
Similar ideas
have been successfully used to describe the  $^3$He+p scattering 
within the hyperspherical
formalism \cite{Fis06}. The success of these approaches suggests 
that the hyperspherical interpolation method
can be applied to heavier systems as well. However, to make it
applicable to other systems, 
a hyperspherical expansion that contains contributions only
from the long-range part of the wave function $\Psi$ should be
introduced.


In this paper, a subset of
hyperspherical basis is constructed for $A$ identical fermions
 that  is expected to be responsible for
long-range behaviour in weakly-bound nuclei and a technique
to calculate matrix elements in such a basis is developed. 
This basis is used to expand the total wave function $\Psi$
into hyperradial and hyperangular parts and then the hyperradial
part is found by solving the standard set
of coupled differential  equations of the hyperspherical functions
method (HSFM).  
The feasibility of the proposed method is studied in the simplest
case of $^5$He. Only those  hyperspherical basis states have been chosen
that correspond to the
$^4$He core  described in the lowest order approximation of the HSFM.
The convergence of such an expansion is investigated
by calculating the  binding energy and r.m.s. radius
of $^5$He with an increasing number of basis states using the Volkov 
effective NN potential V1 \cite{volkov} with two values of
the Majorana parameter, the standard  $m=0.6$ that
corresponds to unbound $^5$He and the 
non-standard  $m=0.3$ that binds this nucleus. For the
latter value, the convergence of the hyperspherical expansion
is also studied for
the overlap integral $\la ^4$He$\otimes n |^5$He$\ra$,
its r.m.s. radius,
spectroscopic factor and asymptotic normalization coefficient (ANC).
The hyperspherical expansion of the present paper should produce
results that are very close to the MCM
 with a closed $0s$-shell $^4$He core described in the oscillator
shell model.
The comparison between both models is presented.

In Sec.II the hyperspherical cluster basis is constructed  and in Sec. III
the expansion onto this basis is discussed. The link
between the hyperspherical cluster model and MCM is shown in Sec. IV,
while the representation of   hyperspherical cluster harmonics by
oscillator cluster wave functions is given in Sec. V. A method to
calculate  matrix elements  in this basis is presented in Sec. VI and
the application to $^5$He is  discussed in Sec. VII. The results obtained
 are summarised in Sec. VIII. Finaly, exact formulae
for the  norm and the two-body NN potential matrix elements as well as
for the overlap integral
are derived in the Appendix.


\section{Hyperspherical cluster harmonics}

A nucleus $A$ can be described in hyperspherical coordinates, that
include the hyperradius $\rho$,
\beq
\rho^2 = \sum_{i=1}^{A-1}\xi_i^2,
\eeqn{rho}
where $\ve{\xi}_i$ are the normalised Jacobi coordinates,
\beq
\ve{\xi}_i =  \sqrt{\frac{i}{i+1}}\left(\frac{1}{i} \sum_{j=1}^i 
\ve{r}_j- \ve{r}_{i+1}\right),
\eeqn{Jacobi}
and a set of $3A-4$ hyperangles $\hat{\ve{\rho}}$. 
The wave function $\Psi$ written in these coordinates can be expanded
into hyperspherical harmonics (HHs) that depend  on the hyperangles.

Let us consider a nucleus $A$ that is   strongly
clusterized as a core $A-1$ plus a weekly-bound
valence nucleon $N$. It is well-known \cite{SSh}
that the spatial HHs for $A$ nucleons can be written
as a product of a spatial HHs for $A-1$ nucleons,
the spherical function $Y_{lm}(\hat{\ve{\xi}}_{A-1})$
and a function $\varphi_n(\theta)$ which is an eigenfunction  of
the angular part of Laplacian in variable $\theta$ defined  
as  $\theta =\arctan  \xi_{A-1}/\rho_c$. Here $\rho_c$ is the
hyperradius for the $A-1$ core, $\rho^2 = \rho_c^2 + \xi_{A-1}^2$.
The angle $\theta$  is associated with the valence nucleon 
and it shows how far  
the valence nucleon is for a fixed size of the core $A-1$.
To describe the long-range behaviour of the last nucleon
for a fixed structure of the core it is necessary to include
as many eigenfunctions $\varphi_n(\theta)$  as possible.

This paper  accounts for  cluster structure $(A-1) + N$ 
of weakly bound
 nuclei by reorganizing known recurrence representation
of HHs in a form that is  similar to the microscopic cluster model.
This is achieved by antisymmetrizing the product of a (known)
completely antisymmetric HH
$Y_{K_c\gamma_c}^{M_{L_c}M_{S_c}M_{T_c}}(\hat{\ve{\rho}}_c)$ 
for the core $A-1$
(spin-isospin part included)  and  a relative angular function
$\varphi_{K_cnlm\sigma\tau} (\theta,\hat{\xi}_{A-1})$
\beq
{\cal Y}_{K_c\gamma_cnl\,LST}^{M_LM_SM_T}(\hat{\ve{\rho}})  = 
{\cal N}_{K_c\gamma_cnl\,LST}^{-1}
\sum_{M_{L_c}M_{S_c}M_{T_c}m\sigma\tau} \eol
\times
(L_cM_{L_c}lm|LM_L)
(S_cM_{S_c}\frac{1}{2}\sigma|SM_S)(T_cM_{T_c}\frac{1}{2}\tau|TM_T)
{\cal A}  
\left(Y_{K_c\gamma_c}^{M_{L_c}M_{S_c}M_{T_c}}(\hat{\ve{\rho}}_c)\,\,
\varphi_{K_cnlm\sigma\tau} 
(\theta,\hat{\xi}_{A-1})
\right).
\eeqn{defY}
Here  $K_c$ is the hypermoment,
$\gamma_c = \{\beta_c L_cS_cT_c\}$, $L_c$, $S_c$, and $T_c$
are the total orbital momentum, spin and isospin
of the core $A-1$ while the $\beta_c$ represents different degenerate harmonics.
Also,
\beq
{\cal A} = \frac{1}{A^{1/2}} \left(1 - \sum_{i=1}^{A-1} P_{iA}\right)
\eeqn{A}
 is the antisymmetrisation operator
which permutes the $A$-th nucleon with nucleons of the core,
${\cal N}_{K_c\gamma_cnl\,LST}$ is a normalization factor, 
 $\hat{\ve{\rho}}_c$ denotes  the hyperangles 
for the $A-1$-body system and
\beq
\varphi_{K_cnlm\sigma \tau} (\theta,  \hat{\ve{\xi}}_{A-1})=  N_{nK_cl} 
\left(\sin\theta\right)^l \left(\cos\theta\right)^{K_c}
P_{n}^{l+1/2,K_c+(3A-8)/2}(\cos 2\theta)
Y_{lm}(\hat{ \ve { \xi}}_{A-1})\chi_{\sigma\tau}(A) 
\eeqn{varpi}
where
 $\cos \theta=\rho_c/\rho $,
$\sin \theta = \xi_{A-1}/\rho $, 
$\cos 2\theta = 1-2\xi^2_{A-1}/\rho^2$, $\chi_{\sigma\tau}(A)$ is
the spin-isospin function of the $a$-th nucleon with the spin and isospin
projection $\sigma$ and $\tau$ respectively, and
\beq
N_{nK_cl}^2 =  
\frac{2n!\,(2n+K_c+l+(3A-5)/2)\Gamma(n+K_c+l+(3A-5)/2)}
{ \Gamma(n+l+3/2) \,\Gamma(n+K_c+(3A-6)/2)}.
\eeqn{nkkl}
 
Since  $Y_{K_c\gamma_c}^{M_{L_c}M_{S_c}M_{T_c}}(\hat{\ve{\rho}}_c)$
and  $\varphi_{K_cnlm\sigma\tau}(\theta,\hat{\ve {\xi}}_{A-1})$ are
 the eigenfunction of the operator of the
kinetic energy in subspaces associated with variables 
$\hat{\ve{\rho}}_c$ and $\{\hat{\ve{\xi}}_{A-1},\theta\}$ and because 
the operator of the
kinetic energy is symmetric and the function
(\ref{defY}) is antisymmetric with respect to any nucleon permutations,
the function ${\cal Y}_{K_c\gamma_cnl\,LST}^{M_LM_SM_T}(\hat{\ve{\rho}})
\equiv {\cal Y}_{K\gamma}(\hat{\ve{\rho}})$
is also an eigenfunction of the $3A-4$-dimensional
angular part  $\Delta_{\hat{\rho}}$ of the Laplacian, 
\beq
\Delta_{\hat{\rho}}{\cal Y}_{K\gamma}(\hat{\ve{\rho}})= - K(K+3A-5)
{\cal Y}_{K\gamma}(\hat{\ve{\rho}}),
\eeqn{f4}
 and therefore it is the HH with the hypermoment
$K = K_c+2n+l$. Here $\gamma = \{K_c\gamma_cnl\,LSTM_LM_SM_T\}$.

Since the HHs
${\cal Y}_{K\gamma}(\hat{\ve{\rho}})$  are
written in the form typical for the cluster wave functions
(\ref{eq1}), they are referred to below as  
 hyperspherical cluster harmonics (HCH). The HCHs 
${\cal Y}_{K\gamma}(\hat{\ve{\rho}})$ with different values of $K$
are orthogonal to each other, however, for the same
$K$ and different $\gamma$ they may be not orthogonal
\beq
\la {\cal Y}_{K'\gamma'}(\hat{\ve{\rho}})|{\cal Y}_{K\gamma}(\hat{\ve{\rho}})\ra  
= \delta _{KK'} {\cal I}_{K \gamma \gamma'}
,
\eeqn{yy}
which is common for different channel functions in 
multichannel cluster models with antisymmetrization.


\section{Expansion of nuclear wave functions
onto hyperspherical cluster basis}


The wave function of a nucleus with one weakly-bound nucleon
can be expanded onto the hyperspherical
cluster basis,
\beq
\Psi =  \rho^{-(3A-4)/2}
\sum_{K\gamma } \chi_{K\gamma}(\rho) {\cal Y}_{K\gamma}(\hat{\ve{\rho}}), 
\eeqn{Psi}
keeping only a restricted number of quantum numbers in the HHs of the
core   and as much relative functions (\ref{varpi})
(or in other words as many $n$'s) as needed to
describe properly  the long-range radial behaviour of the valence nucleon. 
However, when more then one set of quantum numbers for the
HH of the core is present in the expansion (\ref{Psi}),
the hyperspherical cluster basis may not be orthogonal for the same $K$.
The simplest way to deal with this problem is to introduce the
orthogonalised HHs,
\beq
 {\tilde{\cal Y}}_{K {\tilde \gamma}}(\hat{\ve{\rho}}) = \sum_{\gamma}
A_{\gamma {\tilde \gamma}} {\cal Y}_{K\gamma}(\hat{\ve{\rho}}),
\eeqn{ortHH}
which can be achieved, for example,  
by performing the singular value decomposition
of the matrix ${\cal I}_{K \gamma \gamma'}$. Then the decomposition
\beq
\Psi = \rho^{-(3A-4)/2}
\sum_{K{\tilde \gamma }} {\tilde \chi}_{K{\tilde \gamma}}(\rho) 
{\tilde {\cal Y}}_{K{\tilde \gamma}}(\hat{\ve{\rho}})  
\eeqn{Psiort}
onto orthogonal basis 
${\tilde{\cal Y}}_{K {\tilde \gamma}}(\hat{\ve{\rho}})$
leads to a standard set of differential hyperradial equations
of the HSFM,
\beq
\left(\frac{d^2}{d\rho^2}-\frac{{\cal L}_K({\cal L}_K+1)}{\rho^2}
-\frac{2m}{\hbar^2}(E + {\tilde V}_
{K{\tilde \gamma},K{\tilde \gamma}}(\rho))\right)
{\tilde \chi}_{K{\tilde \gamma}}(\rho)
= \frac{2m}{\hbar^2}
\sum_{K'{\tilde \gamma}'\ne K{\tilde \gamma}} 
{\tilde V}_{K{\tilde \gamma},K'{\tilde \gamma'}}(\rho)
{\tilde \chi}_{K'{\tilde \gamma}'}(\rho),
\eeqn{f5}
where ${\cal L}_K = K + (3A-6)/2$, $m$ is the nucleon mass
and  the hyperradial potentials
${\tilde V}_{K{\tilde \gamma},K'{\tilde \gamma}'}(\rho)$ are the  
matrix elements of the  NN interactions
\beq
{\tilde V}_{K{\tilde \gamma},K'{\tilde \gamma}'}(\rho) =
\la {\tilde{\cal Y}}_{K{\tilde \gamma}}(\hat{\ve{\rho}})|
\sum_{i<j}V_{ij}(\ve{r}_i-\ve{r}_j)|
{\tilde{\cal Y}}_{K'{\tilde \gamma}'}(\hat{\ve{\rho}})\ra.
\eeqn{f6}
The functions $\chi_{K\gamma}(\rho)$ corresponding to the non-orthogonal
HCH basis ${\cal Y}_{K\gamma}(\hat{\ve{\rho}})$ with well-defined quantum
numbers $\{K_c\gamma_cnl\}$ are then obtained using the transformation
\beq
\chi_{K\gamma}(\rho) = \sum_{\tilde \gamma}
A_{\gamma {\tilde \gamma}} {\tilde \chi}_{K{\tilde \gamma}}(\rho).
\eeqn{inv}


\section{Link between the   hyperspherical
cluster model and microscopic cluster model} 


If    expansion (\ref{Psi}) retains only a limited number of
  $\{K_c\gamma_c\}$ 
while at each fixed $\{K_c\gamma_c\}$ the summation
over all quantum numbers $n$  is performed, such an expansion
 corresponds to a specific cluster model
of the type (\ref{eq1}). The easiest way to demonstrate this, 
is to retain only
one state $\{K_c\gamma_c\}$ for the core $A-1$ and neglect all spin couplings. 
Then the wave function (\ref{Psi}) can be rewritten as
\beq
\Psi = {\cal A} \left( \sum_n 
{\cal N}_{K_c\gamma_cnl\,LST}^{-1} \,\,
\frac{\chi_{K_c\gamma_cnl}(\rho)}{\rho^{(3A-4)/2}} \,\,
Y_{k_c\gamma_c}(\hat{\ve{\rho}}_c) \,\, \varphi_{K_cnlm\sigma \tau} 
(\theta,\hat{\ve{\xi}}_{A-1})\right).
\eeqn{link1}
If  after the antisymmetrization operator in (\ref{link1}) a closure relation
is inserted, 
\beq
\sum_i |\Psi^{(i)}_{K_c\gamma_c}(\rho_c,\hat{\ve{\rho}}_c)\ra
\la\Psi^{(i)}_{K_c\gamma_c}(\rho_c,\hat{\ve{\rho}}_c)| = 1,
\eeqn{closure}
where $\Psi^{(i)}_{K_c\gamma_c}(\rho_c,\hat{\ve{\rho}}_c)$ is
the $i$-th solution of the Schr\"odinger equation for $A-1$ nucleons
in the hyperspherical basis that keeps only one selected HH, 
$Y_{k_c\gamma_c}(\hat{\ve{\rho}}_c)$,
then the wave function (\ref{Psi}) can be rewritten as
\beq
\Psi = \sum_i {\cal A} \left(\Psi^{(i)}_{K_c\gamma_c}(\rho_c,\hat{\ve{\rho}}_c)
\,\, \phi_{K_c\gamma_c}^{(i)}(\ve{\xi}_{A-1})\right),
\eeqn{link2}
with the relative function $\phi_{K_c\gamma_c}^{(i)}(\ve{\xi}_{A-1})$
determined by the following expression:
\beq
\phi_{K_c\gamma_c}^{(i)}(\ve{\xi}_{A-1}) = \sum_n 
{\cal N}_{K_c\gamma_cnl\,LST}^{-1} \,\left\langle
 \phi_{K_c\gamma_c}^{(i)}(\rho_c,\hat{\ve{\rho}}_c) \,
Y_{k_c\gamma_c}(\hat{\ve{\rho}}_c)\ 
\left| \,\frac{
\chi_{K\gamma}(\sqrt{\rho_c + \xi_{A-1}^2}\,)}{
(\rho_c^2 + \xi_{A-1}^2)^{(3A-4)/4}}
\, \varphi_{K_cnlm\sigma \tau}(\ve{\xi}_{A-1})\right\rangle \right.
\eeqn{relwf}
Thus the HCH expansion of the wave function $\Psi$ is equivalent to 
a microscopic multichannel cluster model that includes 
all the excited states of the core $A-1$
that correspond to the hyperradial  excitations of the lowest state
with $\{K_c\gamma_c\}$.

In the general case, when more sets of the quantum numbers $\{K_c\gamma_c\}$
are retained in the HCH expansion, it is possible to show that the
HCH expansion corresponds to the multichannel cluster model that includes
all the excited states of the core with   total spin $J_c$, that
can be constructed with the values of $L_c$ and $S_c$ retained in
the core description, and
their hyperradial excitations. Such a  derivation is quite cumbersome and 
is not given here.


\section{Link between     hyperspherical
cluster harmonics and oscillator cluster wave functions} 


The HCH ${\cal Y}_{K\gamma}(\hat{\ve{\rho}})$ can be   represented by the cluster
wave functions that contain oscillator shell model wave functions. Such a
representation is used below to develop a technique to calculated the
matrix elements in the HCH basis. The link to the cluster shell model 
can be derived using an expansion of the Jacobi polynomials into the
radial oscillator wave functions $R_{\kappa l}({\xi})$ (see Eqs. 
(\ref{p})-(\ref{R})
of Appendix A) and 
representing the core HH $Y_{K_c\gamma_c}(\hat{\ve{\rho}}_c)$ as
\beq
Y_{K_c\gamma_c}(\hat{\ve{\rho}}_c) = 
\Psi_{0K_c\gamma_c}(\ve{\xi}_1,...,\ve{\xi}_{A-2})/R_{0K_c}
(\rho_c),
\eeqn{corewf}
where $\Psi_{0K_c\gamma_c}(\ve{\xi}_1,...,\ve{\xi}_{A-2})$ is
the translation-invariant oscillator shell model function without
the hyperradial excitations, 
\beq 
R_{0K_c}(\rho_c) = b^{-(3A-6)/2}
\sqrt{\frac{2}{\Gamma(K_c+(3A-6)/2)}}
\left(\frac{\rho_c}{b}\right)^{K_c} e^{-\rho_c^2/2b^2}
\eeqn{rcore}
is the hyperradial oscillator wave function and $b$ is an 
arbitrary oscillator radius. For the lowest possible value
of $K_c$ the $\Psi_{0K_c\gamma_c}(\ve{\xi}_1,...,\ve{\xi}_{A-2})$ is
the usual 0$\hbar \omega$ translation-invariant shell model wave function.
For higher $K_c$, $\Psi_{0K_c\gamma_c}(\ve{\xi}_1,...,\ve{\xi}_{A-2})$ 
can be constructed, for example, using the technique
of Ref. \cite{Tim02}. With Eqs. (\ref{rcore}), (\ref{corewf}) and
(\ref{p})-(\ref{R}) one gets
\beq
{\cal Y}_{K\gamma} (\hat{\ve{\rho}})  =
{\cal N}_{K\gamma}^{-1}
 \sum_{\nu=0}^{n} 
\sum_{\kappa=0}^{\nu} 
B_{nK_cl}^{\nu\kappa} \, \,Z_{\alpha \nu} (\hat{\ve{\rho}})
\eeqn{Y}
where
$\alpha \equiv \{K_c\gamma_c\kappa lLSTM_LM_SM_T\}$,
\beq
Z_{\alpha \nu} (\hat{\ve{\rho}}) =  
b^{K_c+2\nu+l+(3A-3)/2} \rho^{-K_c-l-2\nu}  e^{\rho^2/2b^2} 
{\cal A} 
\Phi^{(b)}_{\alpha}(\ve{\xi}_1,...\ve{\xi}_{A-1})
\eol
\eeqn{Z}
and 
$\Phi^{(b)}_{\alpha}(\ve{\xi}_1,...\ve{\xi}_{A-1})$ 
is a non-antisymmetrized oscillator cluster wave function,
\beq
\Phi^{(b)}_{\alpha}(\ve{\xi}_1,...\ve{\xi}_{A-1})
=
\sum_{M_{L_c}M_{S_c}M_{T_c}m\sigma\tau}
(L_cM_{L_c}lm|LM_L)
(S_cM_{S_c}\frac{1}{2}\sigma|SM_S) (T_cM_{T_c}\frac{1}{2}\tau|TM_T)
\eol
\times
 \Psi_{0K_c\gamma_c}^{M_{L_c}M_{S_c}M_{T_c}}
(\ve{\xi}_1,...,\ve{\xi}_{A-2})\,
R_{\kappa lm}(\ve{\xi}_{A-1})\chi_{\sigma\tau}(A), 
\eeqn{Phi} 
corresponding to the oscillator radius $b$.
The expansion coefficients $B_{nK_cl}^{\nu\kappa}$ are given
by the expression
\beq
B_{nK_cl}^{\nu\kappa}
=(-)^{ \nu+\kappa}\,\, \nu! \,
N_{nK_cl} 
\left(\frac{\Gamma\left(K_c+(3A-6)/2\right)
\Gamma(\kappa+l+3/2)}{4\kappa!}\right)^{1/2}
\eol
\times
 \left( {\begin{array}{c} \nu+l+1/2\\
 \kappa+l+1/2 \end{array} }\right)  
\left( {\begin{array}{c} {n+l+1/2  }\\
 n -\nu \end{array} }\right)
\left( {\begin{array}{c} 
 {n+K_c+l+(3A-7)/2 +\nu }\\
 {n+K_c+l+(3A-7)/2} \end{array} }\right).
\eeqn{B}


\section{Matrix elements in the hyperspherical
cluster basis} 


The matrix elements in the hyperspherical cluster
basis for an arbitrary operator $\hat{O}$ can be calculated
using the link (\ref{Y}) between the HCH and the oscillator cluster 
wave function. This   gives
\beq
\left< {\cal Y}_{K' \gamma'}(\hat{\ve{\rho}})  | \hat{O}|
{\cal Y}_{K \gamma}(\hat{\ve{\rho}})  \right> 
= ({\cal N}_{K' \gamma'}{\cal N}_{K \gamma})^{-1}
\sum_{\nu=0}^n\sum_{\nu'=0}^{n'}
\sum_{\kappa=0}^{\nu} \sum_{\kappa'=0}^{\nu'} 
B_{n'K_c'l'}^{\nu'\kappa'} B_{nK_cl}^{\nu\kappa} 
\left< Z_{\alpha'\nu'}(\hat{\ve{\rho}}) | \hat{O}|
Z_{\alpha \nu}(\hat{\ve{\rho}})   \right> 
\eol
\eeqn{YY}
To calculate the matrix elements $\left< Z_{\alpha'\nu'}(\hat{\ve{\rho}}) 
| \hat{O}|
Z_{\alpha \nu}(\hat{\ve{\rho}})   \right> $
the technique of Ref. \cite{Tim02} can be used that
replaces the integration over hyperangles by the  Laplace transform
of the shell model matrix elements. This results in
\beq
\left< Z_{\alpha'\nu'}(\hat{\ve{\rho}}) | \hat{O}|
Z_{\alpha\nu}(\hat{\ve{\rho}})\right> =
\rho^{-K_c-K_c'-2(\nu+\nu')-l-l'-3A+5} \frac{1}{\pi i} 
\int_{-i\infty}^{i\infty} ds \,e^{s\rho^2} 
s^{-(K_c+K_c'+l+l'+3A-3)/2-\nu-\nu'}
\,O_ {\alpha'} ^{ \alpha}( s^{-1/2})
\eeqn{ZZ}
where
\beq
O_ { \alpha'} ^{ \alpha}(b)
= \int d\ve{\xi}_1...d\ve{\xi}_{A-1}
{\cal A} \left( \Phi^{(b)}_{\alpha'}(\ve{\xi}_1,...\ve{\xi}_{A-1})\right)
\hat{O}{\cal A} \left( \Phi^{(b)}_{\alpha}(\ve{\xi}_1,...\ve{\xi}_{A-1})\right)
\eol
\eeqn{ob}
are the translation-invariant cluster shell model
matrix elements corresponding to the oscillator radius
$b = s^{-2}$.


\subsection{Normalization of hyperspherical cluster harmonics}


The normalization coefficients
${\cal N}_{K\gamma}$ of the HCH  
are obtained using Eqs. (\ref{yy}), (\ref{YY})-(\ref{ob}) with
$\hat{O}=1$. Since the norm 
$I^{SM}_{\alpha\alpha'}= \la {\cal A} \Phi_{\alpha'}|{\cal A}
\Phi_{\alpha}\ra$
of the oscillator shell model cluster functions does not depend on $b$  
the integration over $s$ in Eq. (\ref{ZZ}) can be performed 
 immediately. This gives
\beq
{\cal I}_{K\gamma\gamma'} = \frac{2}{{\cal N}_{K\gamma}
{\cal N}_{K\gamma'}}\sum_{\nu=0}^n\sum_{\nu'=0}^{n}
\sum_{\kappa=0}^{\nu} \sum_{\kappa'=0}^{\nu'}
\frac{\delta_{K_c'+2\kappa'+l',K_c+2\kappa+l}\,\,B_{nK_c'l'}^{\nu'\kappa'} 
B_{nK_cl}^{\nu\kappa} 
I_{ \alpha\alpha'}^{SM}   }
{\Gamma\left(\frac{K_c+K_c'+l+l'+3A-3}{2}+\nu+\nu'\right)}.
\eeqn{I}
Then  ${\cal N}_{K\gamma}$ is found from the condition 
${\cal I}_{K\gamma\gamma} =1$. 

It is useful here to clarify the meaning of 
this normalization coefficient. Let us introduce a 
fractional parentage expansion of the HCH (\ref{defY})
onto the complete set of HHs 
$Y_{K_c\gamma_c}(\hat{\ve{\rho}}_c)$ for the $A-1$ nucleons:
\beq
{\cal Y}_{K\gamma}(\hat{\ve{\rho}}) = 
\sum_{K_c \gamma_cM_{L_c}M_{S_c}M_{T_c}m\sigma\tau}  
\la A K\gamma | A-1 K_c\gamma_c, l\ra \,\,Y_{K_c\gamma_c}(\hat{\ve{\rho}}_c)
\,\varphi_{K_cnlm\sigma\tau} 
(\theta,\hat{\ve{\xi}}_{A-1})
\eol
\times
(L_cM_{L_c}lm|LM_L)
(S_cM_{S_c}\frac{1}{2}\sigma|SM_S)(T_cM_{T_c}\frac{1}{2}\tau|TM_T)
\eeqn{fpe}
where $\la A K\gamma | A-1 K_c\gamma_c, l\ra$ is the fractional expansion
coefficient. Multiplying Eq. (\ref{fpe}) by 
$Y_{K_c\gamma_c}(\hat{\ve{\rho}}_c)$, $\varphi_{K_cnlm\sigma\tau} 
(\theta,\hat{\ve{\xi}}_{A-1})$ and three relevant Clebsch-Gordan coefficients
and then integrating over angular variables and making summations
over all projections of angular momenta and spins, we get
\beq
{\cal N}_{K\gamma}= \sqrt{A} \,\,\, \la A K\gamma | A-1 K_c\gamma_c, l\ra.
\eeqn{N-cfp}
Thus, the normalization coefficient ${\cal N}_{K\gamma}$ 
is directly related to the
fractional expansion coefficient into the  core state on
which the HCH is originally built. The knowledge of this relation is 
very  useful  for calculating the overlap integrals between the wave functions
of nuclei $A$ and $A-1$.

The  overlap $I_{\alpha\alpha'}^{SM}$ that enters Eq. (\ref{I})
can be calculated using the fact that for an arbitrary 
function $\psi_{\alpha} =\phi_{\alpha_1}(1,...,A-1)   \varphi_{\alpha_2}(A) $, 
where $\phi$ is an antisymmetric function,
\beq
\la {\cal A} \psi_{\alpha'} |{\cal A}\psi_{\alpha} \ra 
= A^{1/2} \la \phi_{\alpha_1'}(1,...,A-1)  \varphi_{\alpha_2'} (A  )|{\cal A}\psi_{\alpha} \ra
\eol
=\delta_{\alpha\alpha'}-(A-1)
\la \phi_{\alpha_1'}(1,...,A-1) \varphi_{\alpha_2'}(A)|
\phi_{\alpha_1}(1,...,A-2,A)\varphi_{\alpha_2}(A-1)\ra,
\eeqn{AA}
where ${\cal A}$ is determined by (\ref{A}).
The  exchange term in this expression
can be calculated using
the fractional parentage expansion of the translation-invariant
oscillator wave  function of $A-1$ nucleons. The details of these
calculations and final analytical expressions are given in Appendix B.


\subsection{Matrix elements of central two-body NN interactions}


In this paper, only central two-body NN forces are considered,
\beq
V  = \sum_{i<j}^A \sum_{s,t=0,1} V^{(st)}_{ij}(|\ve{r}_i - \ve{r}_j|)
 \hat{P}_{st}(i,j).
\eeqn{}
Here $\hat{P}_{st}(i,j)$ is the projector into states with spin $s$ and 
isospin $t$ of the pair of nucleons $(i,j)$.
To calculate the matrix elements of this potential
in the HCH basis using Eq. (\ref{YY}), one
needs to know the matrix elements $V^{SM}_{\alpha'\alpha}$
in the oscillator shell model cluster basis,
given by Eq. (\ref{ob}) with $\hat{O} = V$.  
The latter can be found  using  the fact that
for an arbitrary 
function $\psi_{\alpha} =\phi_{\alpha_1}(1,...,A-1)   
\varphi_{\alpha_2}(A) $   
that contains a product of   antisymmetric function $\phi_i$   
\beq
\la {\cal A} \psi_{\alpha'}|V|{\cal A}\psi_{\alpha} \ra 
= A^{1/2} \la \phi_{\alpha_1'}(1,...,A-1)\varphi_{\alpha_2'}(A)|V|{\cal A}
\psi_{\alpha}\ra
=\la \phi_{\alpha_1'}(1,...,A-1) 
\varphi_{\alpha_2'}(A)|V|\phi_{\alpha_1}(1,...,A-1)\varphi_{\alpha_2}(A)\ra 
\eol
- (A-1)\la \phi_{\alpha_1'}(1,...,A-1) 
\varphi_{\alpha_2'}(A)|V|\phi_{\alpha_1}(1,...,A-2,A)
\varphi_{\alpha_2}(A-1)\ra 
.
\eeqn{arb}
Eq. (\ref{arb}) contains direct and exchange terms that can be evaluated
with the fractional parentage expansion technique. The derivation of
formulae for $V^{SM}_{\alpha'\alpha}$ is given in Appendix C and 
the final expression for the potential $V_{\alpha\alpha'}^{SM}$
is
\beq
V_{\alpha\alpha'}^{SM} = 
\sum_{n_0'n_0l_0 st} \frac{1-(-)^{l_0+s+t}}{2}\,\,
{\cal C}^{\kappa\kappa'}_{n_0'n_0l_0 st} \,\, 
\la \psi_{n_0'l_0} || V^{ st}||\psi_{n_0l_0}\ra,
\eeqn{vsm}
where 
\beq
\la \psi_{n_0'l_0} || V^{ st}||\psi_{n_0l_0}\ra
= \int_0^{\infty} dr \, r^2 \, \psi_{n_0'l_0}(r)  \, V^{ st}(\sqrt{2}r)\,
 \psi_{n_0l_0}(r)
\eeqn{me}
is the matrix element of the NN potential
between the two-body oscillator wave functions
$\psi_{n_0l_0}(r)$. The analytical expressions for coefficients 
${\cal C}^{\kappa\kappa'}_{n_0'n_0l_0 st}$ are given in the Appendix C.   

Finally, the hyperradial potentials $V_{K\gamma,K'\gamma'}(\rho)$
 are calculated by using Eq. (\ref{YY}) and applying
the Laplace transform (\ref{ZZ}) to the shell model NN matrix elements
$V_{\alpha\alpha'}^{SM}$. This gives
\beq
 V_{K\gamma,K'\gamma'}(\rho) = 
 ({\cal N}_{K' \gamma'}{\cal N}_{K \gamma})^{-1}
 \sum_{n_0'n_0l_0 st} \sum_{\nu \nu'}
 v^{\nu \nu'}_{n_0'n_0l_0 st}(\rho) 
 w^{\gamma \gamma' \nu \nu'}_{n_0'n_0l_0 st},
\eeqn{finalhrp} 
where 
\beq
v^{\nu \nu'}_{n_0'n_0l_0 st}(\rho) =
\rho^{-K_c-K_c'-2(\nu+\nu')-l-l'-3A+5}
\frac{1}{2\pi i} \int_{-i\infty}^{i\infty} ds \,e^{s\rho^2} s^{-(K_c+K_c'+l+l'+3A-3)/2-\nu-\nu'}
\,\la \psi_{n_0'l_0} || V^{ st}||\psi_{n_0l_0}\ra
\eeqn{vnunup}
and
\beq
w^{\gamma \gamma' \nu \nu'}_{n_0'n_0l_0 st} = 2 \sum_{\kappa=0}^{\nu}
\sum_{ \kappa'=0}^{\nu'}
{\cal C}^{\kappa\kappa'}_{n_0'n_0l_0 st}\,
B_{n'K_c'l'}^{\nu'\kappa'} \,B_{nK_cl}^{\nu\kappa}.
\eeqn{w}


\section{Application to $^5$He}


The hyperspherical cluster model (HCM) proposed here, is tested for
the simplest cluster nucleus, $^5$He = $^4$He + n. 
Only one, the lowest order, hyperspherical
harmonics $Y_{000}(\hat{\rho}_c)$ is retained in the HCH basis states
to describe the core $^4$He so that
the chosen hyperspherical cluster basis
\beq
{\cal Y}_{K\gamma}(\hat{\ve{\rho}}) = 
{\cal N}_{K \gamma}^{-1}  \,
{\cal A} (Y_{000}(\hat{\ve{\rho}}_c) \varphi_{K_cnlm\sigma\tau}
(\theta,\hat{\ve{\xi}}_{A-1})
),
\eeqn{y5h3}
where $n= (K-l)/2$,  is orthogonal. 
The norm of this HCH and the coefficients 
${\cal C}^{\kappa\kappa'}_{n_0'n_0l_0 st}$ required
for calculations of the hyperradial potentials are derived
using the shell model technique  from Appendix C that gives
\beq
I_{\kappa\kappa'}^{SM} = \delta_{\kappa\kappa'} \left(1+4^{-2\kappa-1}
\right)
\eeqn{5heI}
and 
\beq
{\cal C}^{\kappa\kappa'}_{n_0'n_0l_0 st} =
3\left(1+2^{-4\kappa-3}\right)
\delta_{\kappa\kappa'} \delta_{n_0,0} \delta_{n_0',0} \delta_{l_0,0} -
3 \delta_{l_0,0} \left( \delta_{n_0,0} \,
T_{0,0,0,\kappa 1,0 0 \kappa' 1,1}^{0 0 0 0,0 0 n_0' 0} 
+ \delta_{n'_0,0}  \,
T_{0,0,0,\kappa' 1,0 0 \kappa 1,1}^{0 0 0 0,0 0 n_0 0} \right) 
\eol
+
\frac{(2s+1)(2t+1)}{2} \sum_{N\Lambda}
\la 0 0 \,\kappa 1 :1 | \frac{3}{5}| n_0l_0 N\Lambda:1\ra
 \la 0 0 \,\kappa' 1 :1 | \frac{3}{5}| n_0'l_0 N\Lambda:1\ra
\eeqn{ccal} 
To solve the hyperradial coupled equations, the computer code STURMXX
\cite{sturmxx} has been used.

The calculations
are performed with the   Volkov V1 potential   \cite{volkov}
 neglecting the Coulomb interaction. 
The standard value of the Majorana parameter
$m$ = 0.6  used with V1
gives unbound $^5$He. For unbound states, convergence in
total energy may not be achieved and the radius of an unbound nucleus should
diverge. Therefore, keeping in mind future applications for heavier
loosely-bound nuclei, the HCM calculations are also performed for $m$ = 0.3.
With this value of $m$ the NN interaction in odd partial waves 
becomes attractive
leading to a bound $^5$He but keeping the separation energy of the 
valence neutron
relatively small, at  about 2 MeV. 
The resulting binding energies and r.m.s. radii
for both $m= 0.6$ and $m=0.3$
are shown in Fig.1a and Fig.1b  
as a function of the cut-off hypermoment $K_{\max} = 
2n_{\max} + 1$. The binding energy and the r.m.s. radius
of the core $^4$He, calculated in the lowest approximation of the HSFM,
are shown in these figures by horizontal lines (these values do not
depend on $m$). 

\begin{figure}[t]
\centerline{\psfig{figure=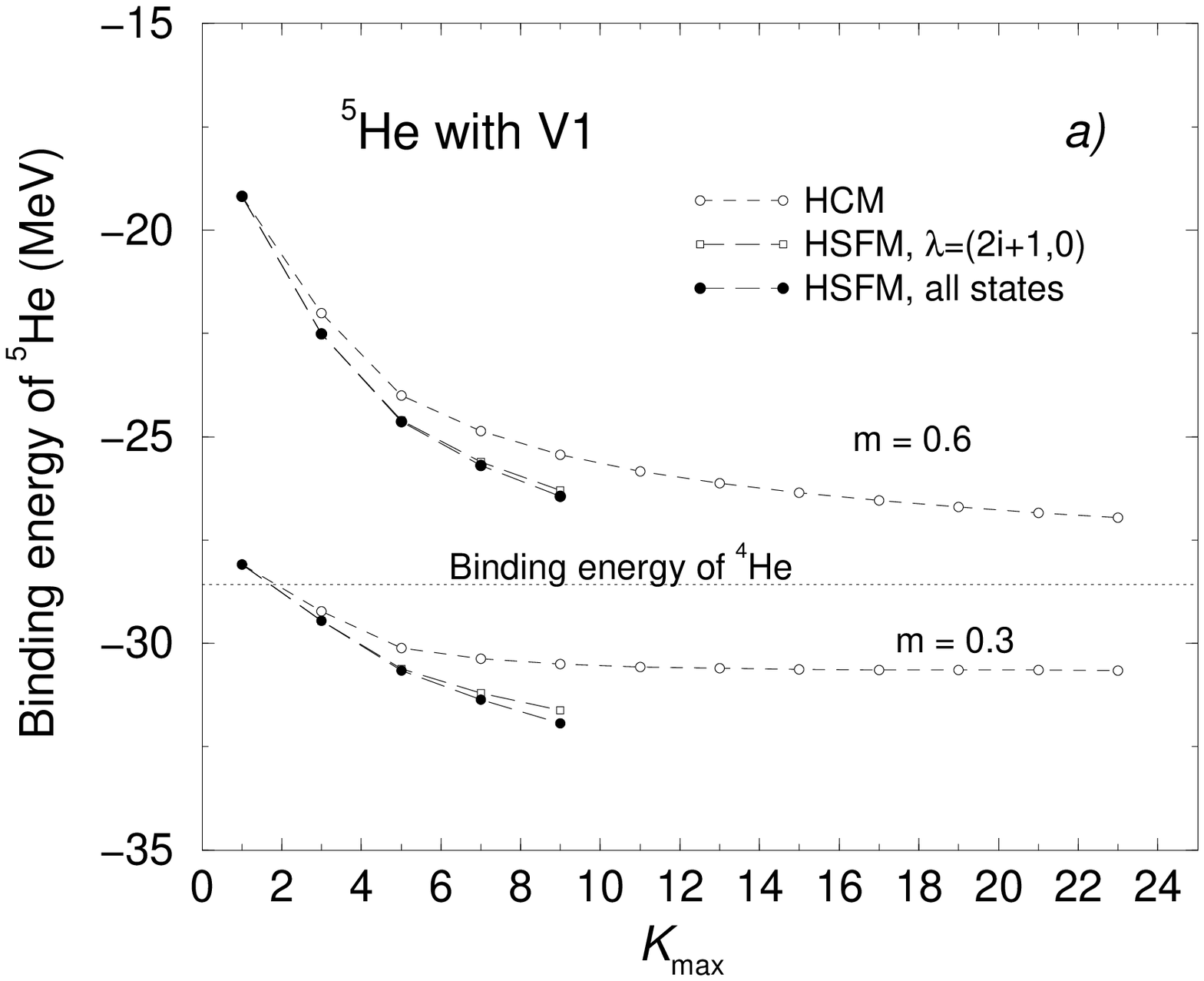,width=0.40\textwidth} 
\hskip 0.5 cm
\psfig{figure=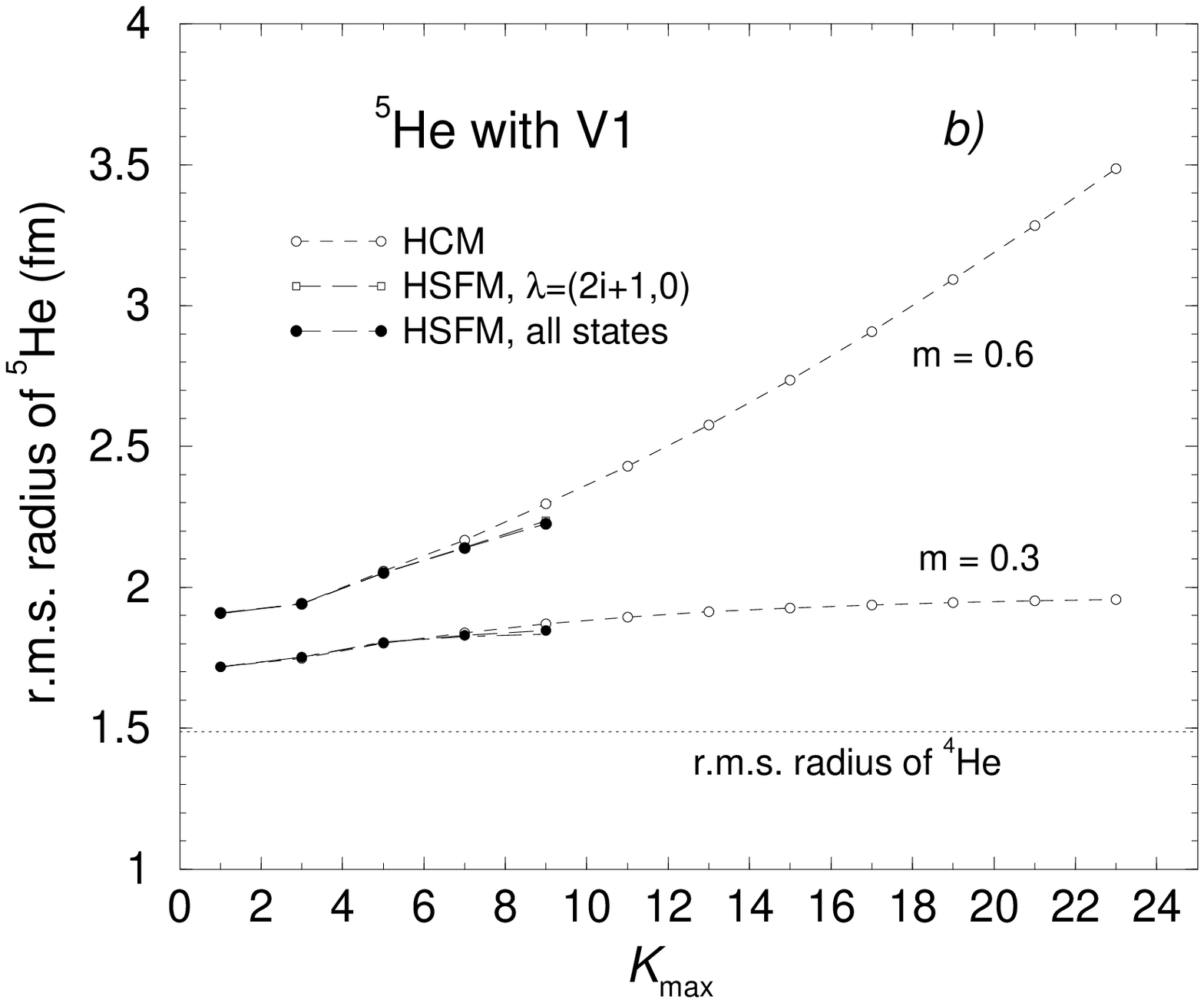,width=0.39\textwidth}}
\caption{Binding energy ($a$) and the r.m.s. radius ($b$)
of $^5$He calculated in the HCM for different cut-off hypermoment
$K_{\max}$. The HSFM calculations, performed in a  restricted basis
that contains only  irreps $\lambda = (2i+1,0)$, where $i$ = 0,1,...,
$(K_{\max}-1)/2$ (open squares), 
and in full basis (filles circles) are shown as well. The binding energy
and the r.m.s. radius of the $^4$He core are represented by dotted
horizontal lines.
}
\end{figure}

One can see that with only twelve 
basis functions the binding energy and radius
of a bound $^5$He corresponding to $m$= 0.3 has practically converged
giving $E$ = -30.66 MeV and $\la r^2\ra^{1/2}$ = 1.96 fm.
For $m$ = 0.6, the binding energy still slowly decreases while the r.m.s.
radius diverges as expected. 
For $K_{\max} \le 9$, the HCM calculations are compared to the 
HSFM calculations performed within the full HH basis
using the technique from Ref. \cite{Tim02}. 
For $K_{\max} = 9$, the difference between 
the HCM and the HSFM is within 1.5 MeV but
the HSFM calculations involve 117 bases states as compared to
 only five states  in the HCM.

It has been shown in \cite{Vas81} that, from the point of view
of the kinematic rotation group $O_{A-1}$,
clustering leads to a  particular symmetry in the total wave function.
According to \cite{Vas81}, 
if each of two clusters have a fixed internal $O_{A-1}$
symmetry the expansion of the total wave function should contain 
only those HHs that correspond
to the irreducible representations (irreps)
($\lambda_1$+$2i$, $\lambda_2$,
$\lambda_3$) of the orthogonal group $O_{A-1}$, where $i$  = 0, 1, 2,
..., $\infty$.
For $^5$He, the HSFM calculations of Ref. \cite{Tim04} 
that include only irreps $\lambda = (2i+1,0)$ give the binding energy
that differs only by about 0.4 MeV from the exact solution. Such calculations
correspond to taking into account all possible   cluster partitions in $^5$He
and they 
are also shown in Fig.1a and Fig.1b. The difference between the binding
energies calculated within the HCM
and the HSFM with  $\lambda = (2i+1,0)$ shows that other cluster
partitions than $^4$He + n  and/or more complicated structure of the 
$^4$He core 
contribute about 1 MeV to the total binding energy of $^5$He. 
On the other hand,
good agreement between the $^5$He radii calculated in HCM and HSFM
indicates that the long-range
behaviour is properly represented in the HCH basis.

To investigate the long-range behaviour in more detail, the 
radial overlap integral $I_{lj}(r)$ between the wave functions of 
bound $^5$He and
$^4$He has been calculated for $m$ = 0.3. 
The definition of this overlap,
\beq
I_{lj}(r) =  \sum_{      m \sigma }  
 (l m \frac{1}{2} \sigma |j m_j )
\int d \hat{\ve{r}} \, Y_{lm}^*(\hat{\ve{r}})\
\, \chi_{\frac{1}{2}\sigma \frac{1}{2}\tau}^{\dagger}
\la \Psi^{^4He}(\ve{x}_1,\ve{x}_2,\ve{x}_3)|
\Psi^{^5He}_{j m_j \tau}(\ve{x}_1,\ve{x}_2,\ve{x}_3,\ve{r})\ra ,
\eeqn{oi} 
includes the wave functions of $^4$He and $^5$He written in
non-normalised Jacobi coordinates $\ve{x}_i$, the last Jacobi coordinate,
$\ve{x}_3 \equiv \ve{r}$, being the distance between the valence nucleon
and the center-of-mass of $^4$He. Such a definition is consistent with
the one used in different reaction theories and in microscopic cluster
models.
The derivation of the final
expression  for this overlap is given in Appendix D. Since the spin-orbital
interaction in these calculations is not present, the total angular
momentum $j$ is ommited below.

The calculated overlap $I_l(r)$
is shown is Fig.2a both in linear and logarithmic scales 
as a function of the cut-off hypermoment $K_{\max}$.  With twelve HCH functions
the overlap $I_l(r)$  practically  converges  for $r \leq 6$ fm. The
r.m.s. radius of this overlap, shown in Fig.3, has almost converged,
giving 3.82 fm.
At large $r$, the $I_l(r)$ should behave as
\beq
\sqrt{5} I_l(r) \rightarrow -i C_l \kappa h_1^{(1)}(i\kappa r)
\,\,\,\, r \rightarrow \infty ,
\eeqn{asym}
where $\kappa = \sqrt{2\mu \epsilon}/\hbar$, $\mu$ is the reduced mass
of $^4$He + n, $\epsilon$ in the neutron separation energy, equal to the
binding energy difference in $^5$He and $^4$He, 
$h_1^{(1)}$ is the Hankel's function of the first kind and
$C_l$ is the asymptotic normalization coefficient (ANC). 
The function $-i\kappa h_1^{(1)}(i\kappa r)$ is shown 
in the inset of Fig. 2a on a logarithmic scale  and one can see that
with increasing $K_{\max}$  the bevaviour of
the overlap $I_l(r)$  approaches the trend
given by (\ref{asym}). 
The ratio $i I_l(r) /\kappa h_1^{(1)}(i\kappa r)$ is shown 
in Fig.2b for $r \leq 12$ fm as a function of $K_{\max}$. Although
this ratio still deviates from constant value for $r > 7$ fm,  the
convergence achived at $5 \leq r \leq 7$ fm with
twelve HCH functions  makes the  unambigous  determination of
$C_l$ possible. The $C_l$  value obtained at these $r$ is  0.72 fm$^{-1/2}$.
The spectroscopic factor for 
the overlap $I_l(r)$, defined as
\beq
S = 5 \int_0^{\infty} dr \, r^2 \, I_l^2(r),
\eeqn{sf}
is shown in Fig.3.
It is almost
independent of $K_{\max}$ and equal to 1.2 which is slightly lower
than the  value of 1.25 obtained in translation-invariant oscillator
0$\hbar \omega$ shell model.

\begin{figure}[t]
\centerline{\psfig{figure=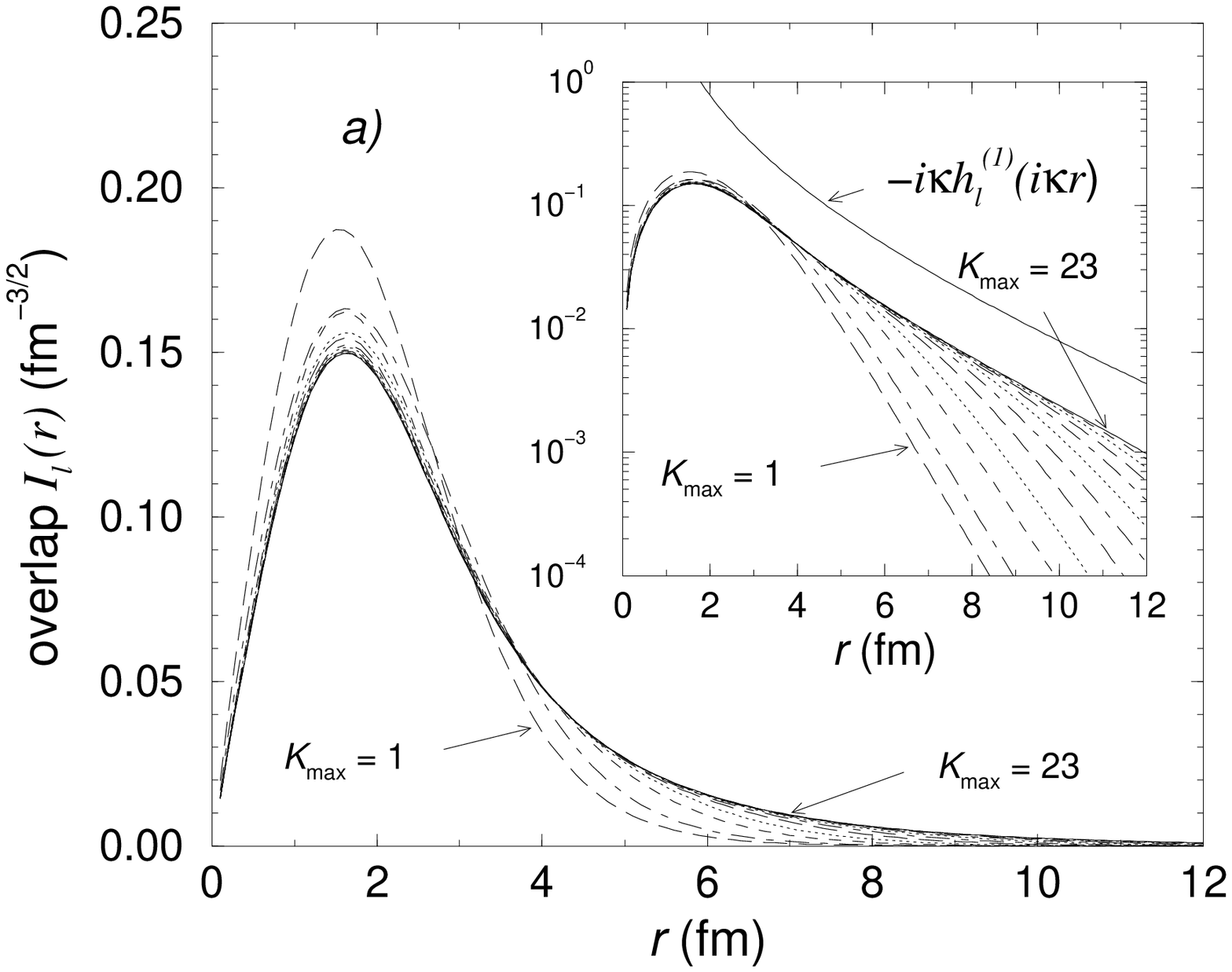,width=0.40\textwidth} 
\hskip 0.5 cm
\psfig{figure=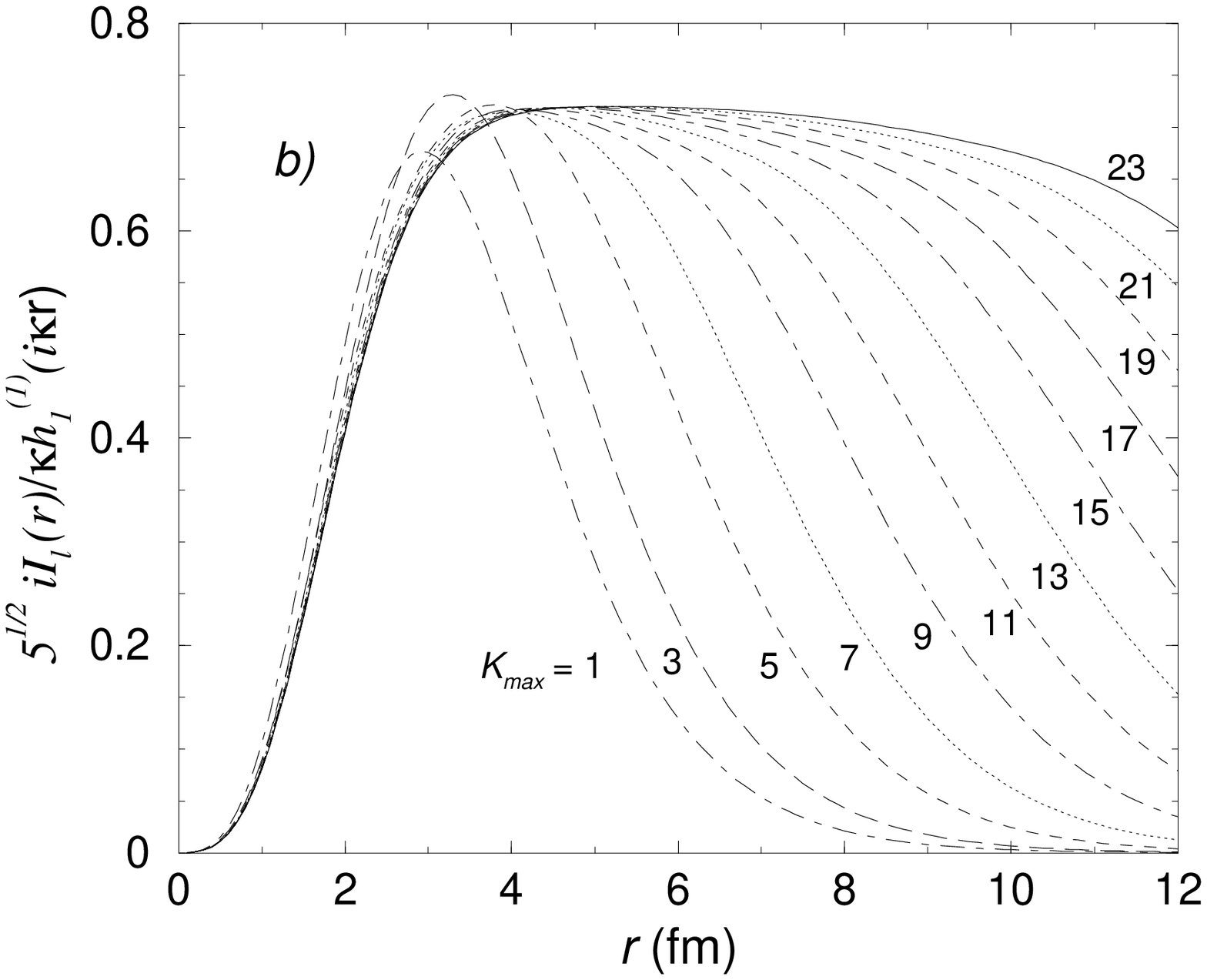,width=0.40\textwidth}}
\caption{Overlap integral  ($a$) for $^5$He, shown in linear (main graph)
and logarithmic (inset) scales, and its ratio to the
  function $-i\kappa h^{(1)}_l(i\kappa r)$ ($b$)
 calculated in the HCM with V1 and $m=0.3$ for different cut-off hypermoment
$K_{\max}$. The function $-i\kappa h^{(1)}_l(i\kappa r)$ is shown
in the inset of ($a$).
}
\end{figure}

The expansion of the $^5$He wave function
onto the HCH basis (\ref{y5h3}) should provide very
similar results to the traditional microscopic cluster model
\beq
\Psi^{^5{\rm He}} = 
{\cal A} (\Psi^{^4{\rm He}}\, \otimes
g_l (r) Y_{lm}(\hat{\ve{r}})\chi_{\frac{1}{2}
\sigma \frac{1}{2} \tau} )
\eeqn{mcm}
with the closed $0s$-shell oscillator wave function for $^4$He. 
In the present paper, the total binding energy of  $^5$He has been calculated
in the MCM as well for different oscillator radii $b$ and the results are 
compared to those calculated in the HCM in Fig.4 for both values
of the Majorana parameter, $m=0.6$ and $m=0.3$. For $m=0.3$, the MCM 
binding energy  
of $^5$He taken at its minumum,
$E$ = $-$29.99 MeV is about 0.7 MeV higher than the HCM result
$E$ = $-$30.66 MeV. This
is because the  $^4$He energy obtained in the lowest order approximation
of the HSFM, $E$ = $-$28.58 MeV is by the same amount lower than the
expectation value  $E$ = $-$27.89 MeV for the $^4$He energy
in the  oscillator 0$s$ shell model basis. 
The neutron separation energy  in both
cases is practically the same, 2.08 MeV in HCM and 2.11 MeV in the MCM.
The overlap integrals calculated in both models 
are also practically the same. The spectroscopic factor 
is 1.2 is both cases and the ANC is 0.73 fm$^{-1/2}$ for MCM as compared to
0.72 fm$^{-1/2}$ for HCM. The calculated functions $g_l(r)$ are also very
similar. The  $g_l(r)$    has asymptotic behaviour 
given by (\ref{asym})
with the same ANC, however it reaches the asymptotic form at
significantly larger radii thus being less convinient to use  for
the ANC determination. 
For $m=0.6$, the results of the MCM and HCM cannot be compared 
directly because
of the different boundary conditions imposed for unbound states in
these models. In Fig. 4 the  resonance energy is presented  
for the MCM, while the energy obtained with $K_{\max} = 23$ is plotted
for the HCM calculations. The HCM result is lower.

\begin{figure}[t]
\centerline{\psfig{figure=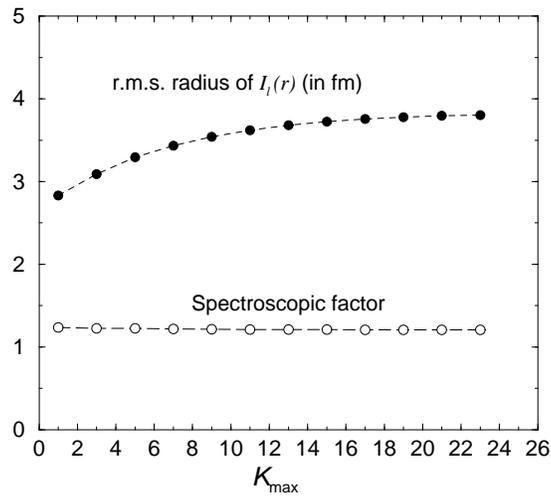,width=0.40\textwidth} }
\caption{The r.m.s. radius and the spectroscopic factor of the
overlap integral $I_l(r)$
for $^5$He calculated in the HCM for different cut-off hypermoment
$K_{\max}$.
}
\end{figure}

\begin{figure}[t]
\centerline{\psfig{figure=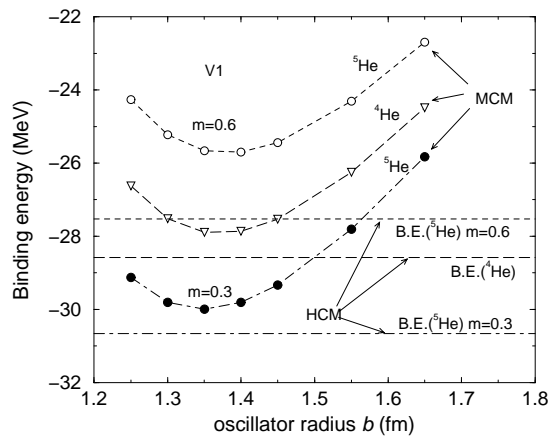,width=0.40\textwidth} }
\caption{Binding energy  
of $^5$He calculated in the MCM  for various oscillator radii $b$
in comparison with HCM values.
}
\end{figure}

\section {Summary and conclusions}

This work has demonstrated that the HH basis can be reorganised
in such a way as to represent the cluster structure of weakly-bound nuclei
and the long-range behaviour of their valence nucleons. 
This can be achieved by antisymmetrizing the product of
completely antisymmetric HH for the core and the relative
hyperangular function  of the last nucleon.
The hyperspherical cluster model that retains only a few
HHs for the core while performing summation over all quantum numbers
related to  the  valence nucleon strongly resembles traditional
microscopic cluster models. However, in the HCM the factorization of
the wave function into the core and  the relative wave functions is absent 
which reflects the influence of the
last nucleon on the core wave function.    
Such influence is absent in microscopic cluster models where
the wave function of the core does not depend on the position of the
last nucleon. The factorization into core and relative wave functions
can be achieved in HCM by projection onto eigenfunctions of the
$A-1$-body Hamiltonian. This makes
the HCM to be equivalent to a multichannel
microscopic cluster model with more complicated core structure
and with all possible   
  monopole core excitations built on HHs retained in the
HCH expansion. On the other hand, the HCM is a narrower version of the HSFM
because its hyperradial part 
is found by solving the same coupled set
of differential equations that arises in the traditional HSFM.

The feasibility of the HCM calculations has been studied using $^5$He
as an  example. With only one HH for the core $^4$He
retained in the HCH expansion,
such a model should be very close to the microscopic cluster model with
the core described by the lowest approximation of the HSFM
because all the monopole hyperradial excitations lie too high in energy.
The calculations, performed with a simple soft-core V1 potential,
have confirmed that it is  possible to get a converged solution for
binding energy, r.m.s. radius, overlap integral and ANC with a small
number of basis functions if the strength of the Majorana  force
is modified to bound $^5$He. The values obtained of these quantities
are very close to those calculated
in the traditional MCM with the oscilator radius that gives the
minimum energy in $^4$He, as expected. 
For unbound $^5$He, the convergence of the HCH expansion can not
be achieved and the r.m.s. diverges because of the wrong boundary conditions
imposed on the HCM wave function.
Scattering problem should be solved in this case rather
than a bound state problem.


The long-range behaviour of the last (bound) nucleon, governed by the
 Hankel  function,
is well reproduced when the number of   HCHs increases.
In the present short study, twelve HCH basis states are enough to derive
the ANC although more functions are needed
to reproduce the exponential decrease at $r > 7$ exactly. However, a
further increase of basis states in the framework of HCM may not be
the best option.
The HCH basis can be used in the hyperspherical interpolation
approach of Refs. \cite{BZh} and \cite{G1} where  asymptotic
conditions are incorporated into the system of radial equations, making
the size of the basis smaller. Scattering problem can
be considered in this basis as well.

In the present test study, only the simplest
core structure has been considered. Application to a case with
a more complicated core,  for example,  the one with
particle-hole excitations, can be done straightforwardly using
formulae from Sec. VI and Appendix C. Currently, the only missing input 
quantities for such calculations are the FPCs. However, they
can either derived using the recursive procedure of Ref. \cite{BN} or
by overlaping the HHs obtained in the shell model approach  developed
in Ref. \cite{Tim02}. 

Finally, the ideas presented in this paper can be extended to nuclei with two 
loosely-bound
valence nucleons, and in particular to Borromean nuclei. This would
include more complicated matrix elements and fractional parentage expansions
but will make it possible to achieve a proper  three-body description
of such systems at large distances within a many-body formalism.

\section*{Acknowledgements}
I am grateful to   I.J. Thompson for 
providing me with his computer code STURMXX and to 
P. Descouvemont
for his help with MCM calculations. The EPSRC grant GR/T28577/01 is
also acknowledged for its partial support.

\section{appendix}

\subsection{Expansion of Jacobi polynomials into oscillator wave functions}

The Jacobi polynomial $P_n^{\alpha,\beta}(\cos 2\theta)$
can be represented as follows:
\beq
P_n^{\alpha,\beta}(1-2\xi^2/\rho^2) = \sum_{\nu=0}^{n} (-)^{\nu}
 \left( {\begin{array}{c} {n+\alpha }\\
 n-\nu \end{array} }\right)
\left( {\begin{array}{c} {n+\alpha+\beta+\nu }\\
 {n+\alpha+\beta} \end{array} }\right)
\frac{\xi^{2\nu}}{\rho^{2\nu}}
.
\eeqn{p}
On the other hand,
\beq
{\xi}^{l+2\nu}e^{-\xi^2/2b^2} = 
b^{2\nu+l+3/2}\, 
\nu !
\sum_{\kappa=0}^{\nu}(-1)^{\kappa} 
\left( {\begin{array}{c} \nu+l+1/2\\
 \kappa+l+1/2 \end{array} }\right)
 \sqrt{\frac{\Gamma(\kappa+l+\frac{3}{2})}{2\kappa!}}
R_{\kappa l}({\xi}),
\eeqn{fx}
where
\beq
R_{\kappa l}({\xi}) = b^{-3/2} \sqrt{\frac{2\kappa!}
{\Gamma(\kappa+l+\frac{3}{2})}}
\left(
\frac{\xi}{b}\right)^l 
e^{-\xi^2/2b^2} L_{\kappa}^{l+1/2}\left(\xi^2/b^2\right)
\eeqn{R}
is the radial oscillator wave function with an arbitrary oscillator radius $b$
and $L_{\kappa}^{l+1/2}$ is the generalised Laguerre polynomial.
Eq. (\ref{fx}) can be proven by induction. Inserting (\ref{fx}) and (\ref{R})
into (\ref{p}) gives relation between Jacobi polynomials
and the oscillator wave functions.

\subsection{Normalization of hyperspherical cluster harmonics}

To calculate the exchange term in Eq. (\ref{AA}) the fractional parentage
expansion of the oscillator wave  function of the core can be used 
both in bra- and ket-vectors. This
reads:
 
\beq
\Psi_{0K_c\gamma_c}^{M_{L_c}M_{S_c}M_{T_c}}(\ve{\xi}_1,...,\ve{\xi}_{A-2}) =
\sum_{N''\gamma''\kappa_1l_1} \la A-1 N_c K_c\gamma_c|A-2 N''K''\gamma'',\kappa_1l_1\ra
\sum_{M_L''M_S''M_T''m_1\sigma_1\tau_1} |A-2 N''K''\gamma'' M_L''M_S''M_T''\ra  
\eol
\times
(L''M_L''l_1m_1|L_cM_{L_c})
(S''M_S''\frac{1}{2}\sigma_1|S_cM_{S_c}) (T''M_T''\frac{1}{2}\tau_1|T_cM_{T_c}) 
\psi_{\kappa_1 l_1m_1}(\ve{\xi}_{A-2})\chi_{\sigma_1\tau_1}(A-1)
\eol
\eeqn{fpc}
where $|A-2 N''K''\gamma'' M_L''M_S''M_T''\ra  $ is a translation-invariant
shell model wave function for $A-2$ nucleons
with total number of oscillator quanta equal to $N''$,
well-defined hypermoment $K''$ and other quantum numbers
denoted by $\gamma''$. The fractional parentage coefficients
(FPCs)
$\la A-1 N_cK_c\gamma_c|A-2 N''K''\gamma'',\kappa_1l_1\ra$  
for this expansion have been introduced in Ref. \cite{Kur71}. 
In Eq. (\ref{fpc}) $N_c$ is the total number of oscillator quantum and
$N_c = K_c = K''+2\kappa_1+l_1$. These FPCs
can be calculated using the ideas of Ref. \cite{BN}. Alternatively,
they can be derived by explicit
overlaping   shell model wave functions with well-defined hypermoment,
constructed in Ref. \cite{Tim02}, for $A-1$ and $A-2$ nucleons.
It could even be more practical to expand the wave function
$\Psi_{0K_c\gamma_c}^{M_{L_c}M_{S_c}M_{T_c}}(\ve{\xi}_1,...,\ve{\xi}_{A-2}) $
into wave functions for $A-2$ without a well-defined  value of hypermoment.
Therefore, below, the index $K''$ in the wave functions
for $A-2$ nucleons in all FPCs will be omitted.
The index $N_c$ will be omitted
as well because $N_c = K_c$.

The wave function of the two last nucleons
in the bra-vector of the exchange term (\ref{AA})
is expressed in coordinates $\ve{\xi}_{A-2}$ and $\ve{\xi}_{A-1}$ while
  in the ket-vertor it is expressed in $\ve{\xi}'_{A-2}$ and $\ve{\xi}'_{A-1}$,
\beq
\ve{\xi}_{A-2}' = \sqrt{\frac{A-2}{A-1}} 
\left(\frac{1}{A-2}\sum_{i=1}^{A-2} \ve{r}_i-\ve{r}_A\right),
\eol 
 \ve{\xi}_{A-1}' = \sqrt{\frac{A-1}{A}} 
\left(\frac{1}{A-1}\left(\sum_{i=1}^{A-2} \ve{r}_i+\ve{r}_A\right)
-\ve{r}_{A-1}\right)
\eeqn{xi}  
Using the Talmi-Moshinsky transformation 
\beq
|\psi_{\kappa_1 l_1}(\ve{\xi}_{A-2})\psi_{\kappa l}(\ve{\xi}_{A-1}):L_0M_0\ra =
\sum_{\kappa_1 l_1\kappa_1' l_1'}
\la \kappa'l'\kappa_1'l_1':L_0|A(A-2)| \kappa_1l_1\kappa l:L_0\ra
|\psi_{\kappa' l'}(\ve{\xi}_{A-1}')\psi_{\kappa_1' l_1'}(\ve{\xi}_{A-2}'):L_0M_0\ra
\eeqn{talmi1}
where 
$\la \kappa'l'\kappa_1'l_1':L_0|A(A-2)| \kappa_1l_1\kappa l:L_0\ra
$ is the
Talmi-Moshinsky coefficient (the ordering and meaning of symbols in this 
coefficient is the same as in Ref. \cite{Trl}) 
and $2\kappa_1+l_1+2\kappa+l = 2\kappa_1'+l_1'+2\kappa'+l'$, 
we get 
 for the overlap
$I_{\alpha\alpha'}^{SM}$ the following expression:
\beq
I_{\alpha\alpha'}^{SM} = \la {\cal A} 
 \Phi^{(b)}_{\alpha'} 
| {\cal A} \Phi^{(b)}_{\alpha} 
\ra =
\delta_{\alpha\alpha'}- \delta_{LL'}\delta_{SS'}\delta_{TT'}(A-1) 
\sum_{N''\gamma''}
U(S''\frac{1}{2}\frac{1}{2}S;S_cS_c')\,
U(T''\frac{1}{2}\frac{1}{2}T;T_cT_c')
\eol \times
 (-)^{S+S''+S_c+S_c'+T+T''+T_c+T_c'} 
 \sum_{\kappa_1l_1\kappa_1'l_1'} \la A-1 K_c\gamma_c|A-2 N''\gamma'',\kappa_1l_1\ra
 \la A-1 K_c'\gamma_c'|A-2 N''\gamma'',\kappa_1'l_1'\ra
\eol \times 
\sum_{L_0} (-)^{l'+l_1'-L_0}U(L''l_1Ll;L_cL_0)\,U(L''l_1'Ll';L_c'L_0)
 \la \kappa'l'\kappa_1'l_1':L_0|A(A-2)| \kappa_1l_1\kappa l:L_0\ra
\eol
\eeqn{}
where $ U(j_1j_2Jj_3;j_{12}j_{23}) = \hat{j}_{12}\hat{j}_{23}
W(j_1j_2Jj_3;j_{12}j_{23})$, $\hat{j} = \sqrt{2j+1}$ 
and $W$ is the Racah coefficient.

\subsection{Matrix elements of central two-body NN interactions}

The matrix elements 
$ \la {\cal A} \psi_{\alpha'}|V|{\cal A}\psi_{\alpha} \ra 
$
between the antisymmetrized oscillator cluster
shell model wave functions contain direct and exchange terms,
\beq
\la {\cal A} \psi_{\alpha'}|V|{\cal A}\psi_{\alpha} \ra 
= V_{dir}-(A-1)V_{ex}.
\eeqn{arb1}
The direct term 
\beq 
V_{dir} =\la \phi_{\alpha_1'}(1,...,A-1) 
\varphi_{\alpha_2'}(A)|V|\phi_{\alpha_1}(1,...,A-1)\varphi_{\alpha_2}(A)\ra
\eeqn{dir}
can be separated into two parts,
\beq
V_{dir} = V_{dir}^{(1)} + V_{dir}^{(2)}  = \delta_{\alpha_2, \alpha_2'}
\la \phi_{\alpha_1'}(1,...,A-1)|\sum_{i<j}^{A-1}V_{ij}|\phi_{\alpha_1}(1,...,A-1)\ra 
\eol
 +
\la \phi_{\alpha_1'}(1,...,A-1) \varphi_{\alpha_2'}(A)|\sum_{i=1}^{A-1}V_{iA}|
\phi_{\alpha_1}(1,...,A-1)\varphi_{\alpha_2}(A)\ra, 
\eeqn{}
the first of which is the  expectation value of the NN potential 
for $A-1$ nucleons of the core in the standard translation-invariant
shell model  basis,                               
\beq
V_{dir}^{(1)} = \delta_{\kappa'\kappa}\delta_{ll'}\delta_{mm'} \,
\la \Psi_{0K_0'\gamma_0'}^{M'_{L_0}M'_{S_0}M'_{T_0}}(\ve{\xi}_1,...,\ve{\xi}_{A-2})
|\sum_{i<j}^{A-1}V_{ij}|
\Psi_{0K_0\gamma_0}^{M_{L_0}M_{S_0}M_{T_0}}(\ve{\xi}_1,...,\ve{\xi}_{A-2})\ra,
\eeqn{vdir}
and the second term $V_{dir}^{(2)}$ being   the
 folding potential between the last nucleon and the core. 
The exchange potential, 
\beq
V_{ex} =
\la \phi_{\alpha_1'}(1,...,A) 
\varphi_{\alpha_2'}(A-1)|V|\phi_{\alpha_1}(1,...,A-2,A-1),
\varphi_{\alpha_2}(A)\ra
\eeqn{ex} 
can be represented by four terms,
\beq
V_{ex} = V_{ex}^{(1)} + V_{ex}^{(2)} + V_{ex}^{(3)} + V_{ex}^{(4)},  
\eeqn{vex}
according to the following separation of the two-body interaction potential:
\beq
\sum_{i<j}^A V_{ij} = V_{A-1,A} + \sum_{i=1}^{A-2} V_{iA}+ 
\sum_{i=1}^{A-2} V_{iA-1} +\sum_{i<j}^{A-2} V_{ij}.
\eeqn{}

The terms $V_{dir}^{(2)}$ and $V_{ex}^{(1)}$ resemble each other. They  can be
easily calculated by separating the
wave function of the  $(A-1,A)$  pair 
in the cluster wave function
 $\Phi^{(b)}_{\alpha}$.
This can be achieved by using the fractional parentage expansion (\ref{fpc})
of the core wave function 
$\Psi_{0K_c\gamma_c}^{M_{L_c}M_{S_c}M_{T_c}}(\ve{\xi}_1,...,\ve{\xi}_{A-2})$
combined  with the Talmi-Moshinsky
 transformation:
\beq
|\psi_{\kappa_1 l_1}(\ve{\xi}_{A-2})\psi_{\kappa l}(\ve{\xi}_{A-1}):L_0M_0\ra =
\sum_{n_0l_0 N\Lambda}
\la \kappa_1 l_1\kappa_1 l_1 :L_0|\frac{A-2}{A}| n_0l_0N\Lambda:L_0\ra \,
|\psi_{n_0l_0}(\ve{\zeta}_1)\psi_{N\Lambda}(\ve{\zeta}_2):L_0M_0\ra,
\eeqn{talmi2}
in which
\beq
\ve{\zeta}_1 = \frac{1}{\sqrt{2}}(\ve{r}_{A-1}-\ve{r}_A), \,\,\,\,\,\,\,\,\,\,\,\,\,
\ve{\zeta}_2 = \sqrt{\frac{2(A-2)}{A}} \left(\frac{1}{A-2}\sum_{i=1}^{A-2}\ve{r}_i
-\frac{\ve{r}_{A-1}+\ve{r}_A}{2}\right).
\eeqn{zeta}
Then the  wave function $\Phi^{(b)}_{\alpha}$ reads as follows
\beq
\Phi^{(b)}_{K_c\gamma_c \kappa lLSTM_LM_SM_T}(\ve{\xi}_1,...\ve{\xi}_{A-1})  =
\sum_{N''\gamma'' \kappa_1 l_1    M_L''M_S''M_T''
 } |A-2 N''\gamma'' M_L''M_S''M_T''\ra \,
\la A-1 K_c \gamma_c |A-2 N''\gamma'' , \kappa_1l_1\ra
\eol \times
\sum_{ n_0l_0 N\Lambda L_0  }
\la \kappa_1l_1,\kappa l:L_0|\frac{A-2}{A}|n_0l_0, N\Lambda : L_0\ra \,\,
|\psi_{n_0l_0}(\ve{\zeta}_1),\psi_{N\Lambda}(\ve{\zeta}_2):L_0M_{L_0}\ra \,
U(L''l_1Ll;L_cL_0) \eol \times
\sum_{ S_0T_0  }
\chi_{S_0M_{S_0}T_0M_{T_0}}(A-1,A)\,
U(S''\frac{1}{2}S\frac{1}{2};S_cS_0) U(T''\frac{1}{2}T\frac{1}{2};T_cT_0)
 \eol
\times \sum_{M_{L_0}M_{S_0}M_{T_0 }}(L''M_L''L_0M_{L_0}|LM_L)
(S''M_S''S_0M_{S_0}|SM_S)(T''M_T''T_0M_{T_0}|TM_T).
\eol
\eeqn{}
The expansion of the wave function $\Phi^{(b)}_{\alpha'}$ in the
bra vector looks exactly the same apart from an  additional phase
factor $(-)^{l_0+S_0+T_0}$ which appears due to interchange of
$A-1$-th and $A$-th nucleons. As the result, we get:
\beq
V_{dir}^{(2)}-(A-1)V_{ex}^{(1)} =
(A-1)
\la \Phi^{(b)}_{\alpha'}(1,...,A-1,A) 
- \Phi^{(b)}_{\alpha'}(1,...,A,A-1)
|V_{A-1,A}|
\Phi^{(b)}_{\alpha}(1,...,A-1,A)\ra 
\eol
= (A-1) 
\sum_{n_0'n_0l_0  st}
(1-(-1)^{l_0+s+t})\,\, C_{n_0'n_0l_0st}^{\alpha'\alpha}
\,\, \la \psi_{n_0'l_0}||V^{(st)} || \psi_{n_0l_0}\ra, 
\eeqn{}
where
 \beq
C_{n_0'n_0l_0st}^{\alpha'\alpha}
=
\sum_{N''\gamma'' \kappa_1 l_1 \kappa_1'l_1'}
\la A-1 K_c \gamma_c |A-2 N''\gamma'' , \kappa_1l_1\ra \,
\la A-1 K_c' \gamma_c' |A-2 N''\gamma'' , \kappa_1'l_1'\ra \,
U(S''\frac{1}{2}S\frac{1}{2};S_cs) 
\eol \times
U(T''\frac{1}{2}T\frac{1}{2};T_ct) \,
U(S''\frac{1}{2}S\frac{1}{2};S_c's) \,U(T''\frac{1}{2}T\frac{1}{2};T_c't)
\sum_{   L_0  }
U(L''l_1Ll;L_cL_0)
\,U(L''l_1'Ll';L_c'L_0)
\eol
\times 
\sum_{N\Lambda  }
\la \kappa_1l_1,\kappa l:L_0|\frac{A-2}{A}|n_0l_0, N\Lambda  :L_0\ra \,
\la \kappa_1'l_1',\kappa'l':L_0|\frac{A-2}{A}|n_0'l_0, N\Lambda  :L_0\ra
\eol
\eeqn{}

To calculate the exchange term   $V_{ex}^{(2)}$ 
\beq
V_{ex}^{(2)}  = 
\la \phi_{\alpha_1'} (1,...,A) \varphi_{\alpha_2'} (A-1)|\sum_{i=1}^{A-2}V_{iA}|
\phi_{\alpha_1} (1,...,A-2,A-1)\varphi_{\alpha_2} (A)\ra  
\eol
=(A-2)(\la \phi_{\alpha_1'} (1,...,A-2,A) \varphi_{\alpha_2'} (A-1)|V_{A-2,A}|
\phi_{\alpha_1} (1,...,A-2,A-1)\varphi_{\alpha_2}(A )\ra 
\eeqn{ex2}
it is convenient to separate the wave function 
of the pair $(A-2,A-1)$ from the
cluster function
$\Phi^{(b)}_{\alpha}$ using a two-nucleon 
frantional parentage expansion for the core,
\beq
\Psi_{0K_c\gamma_c}^{M_{L_c}M_{S_c}M_{T_c}}(\ve{\xi}_1,...,\ve{\xi}_{A-2}) =
\sum_{N''K''\gamma'' N \Lambda n_0 l_0L_0 S_0 T_0} 
\la A-1K_c\gamma_c| A-3 N''K''\gamma'';N\Lambda,n_0l_0S_0T_0(L_0):
L_cS_cT_c\ra
 \eol
\times
\sum_{  M_L'' M_S''M_T'' M_{L_0} M_{S_0} M_{T_0}}
| A-3 N''K''L''M_L''S''M_S''T''M_T'' \ra
|\psi_{N\Lambda}(\ve{\eta}_2), \psi_{n_0l_0}(\ve{\eta}_1):L_0M_{L_0}\ra
(L''M_L''L_0M_{L_0}|L_cM_{L_c}) 
 \eol
\times
(S''M_S''S_0M_{S_0}|S_cM_{S_c}) 
(T''M_T''T_0M_{T_0}|T_cM_{T_c})\chi_{S_0M_{S_0}T_0M_{T_0}}(A-2,A) ,
\eeqn{fpc2}
where $| A-3 N''K''L''M_L''S''M_S''T''M_T'' \ra$ is the
translation-invariant shell model wave function for
$A-3$ nucleons with $N''$ quanta, hypermoment $K''$ and
other quantum numbers denoted by $\gamma''$ and
$\la A-1K_c\gamma_c| A-3 N''K''\gamma'';N\Lambda,n_0l_0S_0T_0(L_0):
L_cS_cT_c\ra$ is the two-nucleon FPC. As explained in the previous section,
$K''$ is omitted below. In Eq. (\ref{fpc2}), the
coordinates $\ve{\eta}_1$ and $\ve{\eta}_2$ are the following,
\beq
\ve{\eta}_1 = \frac{1}{\sqrt{2}}\left(\ve{r}_{A-2}-\ve{r}_{A-1}\right), \,\,\,\,\,\,\,
\ve{\eta}_2 = \sqrt{\frac{2(A-3)}{A-1}}\left(\frac{1}{A-3}\sum_{i=1}^{A-3}\ve{r}_i 
- \frac{\ve{r}_{A-2}+\ve{r}_{A-1}}{2}\right).
\eeqn{eta}
Exchange of nucleons $A-1$ and $A$ leads to new coordinates,
$\ve{\eta}_1'$ and $\ve{\eta}_2'$, that are obtained from 
$\ve{\eta}_1$ and $\ve{\eta}_2$
by replacing $\ve{r}_{A-1}$ by $\ve{r}_A$.
Applying the expansion
(\ref{fpc2}) and  using the transformation
\beq
\sum_{M_{L_0}M_{L_c}m} (L''M_L''L_0M_{L_0}|L_cM_{L_c})(L_cM_{L_c}lm | LM)
|\psi_{N\Lambda}(\ve{\eta}_2), \psi_{n_0l_0}(\ve{\eta}_1):L_0M_{L_0}\ra \psi_{\kappa lm}
(\ve{\xi}_{A-1}) = 
\eol
= \sum_{L_0''M_{L_0}''L_c''M_{L_c}'' \kappa''l''m''N'\Lambda'n_0'l_0'}
(L''M_L''L_0''M_{L_0}''|L_c''M_{L_c}'')(L_c''M_{L_c}''l''m'' | LM)
\eol \times
|\psi_{N'\Lambda'}(\ve{\eta}_2'), \psi_{n_0'l_0'}(\ve{\eta}_1'):L_0''M_{L_0}''\ra \psi_{\kappa'' l''m'}
(\ve{\xi}_{A-1}')
T^{N\Lambda n_0l_0,N'\Lambda'n_0'l_0'}_{L'',L_0L_c\kappa l,
L_0''L_c''\kappa''l'',L},
\eeqn{Talmi2} 
the wave function
$\Phi^{(b)}_{K_c\gamma_c \kappa lLSTM_LM_SM_T}(1,...,A-2,A-1,A)$ 
can be rewritten as
\beq
\Phi^{(b)}_{K_c\gamma_c \kappa lLSTM_LM_SM_T}(1,...,A-2,A-1,A) =
\sum_{\begin{array}{c}{\scriptstyle
N''\gamma'' N\Lambda N'\Lambda' n_0l_0n_0'l_0' n''l''} \\ 
{\scriptstyle   L_0  L_0' L_c'' S_0  S_0' S_c'' T_0  T_0' T_c''
\{\mu\} }\end{array}} 
|A-3 \,N''\gamma''M_L''M_S''M_T''\ra
\eol
\la A-1K_c\gamma_c| A-3 N''\gamma'';N\Lambda,n_0l_0S_0T_0(L_0):
L_cS_cT_c\ra\,
|\psi_{N'\Lambda'}(\ve{\eta}_2')
\psi_{n_0'l_0'}(\ve{\eta}_1'):L_0'M_{L_0}'\ra\varphi_{\kappa''l''m''}(\ve{\xi}_{A-1}')
\eol \times
T^{N\Lambda n_0l_0,N'\Lambda'n_0'l_0'}_{L'',L_0L_c\kappa l,L_0''L_c''\kappa''l'',L}
(L''M_L''L_0'M_{L_0}'| L_c''M_{L_c}'' ) (L_c''M_{L_c}'' l''m''| LM_L)
\eol \times 
\chi_{S_0'M_{S_0}'T_0'M_{T_0}'}(A-2,A)
\chi_{\sigma'\tau'}(A-1) \,\,
\hat{S}_c\hat{S}_c''\hat{S}_0\hat{S}_0'
\hat{T}_c\hat{T}_c''\hat{T}_0\hat{T}_0'
\left\{ \begin{array}{lll} {S} &{ \frac{1}{2} }&{S_c''}\\
{\frac{1}{2}}&{\frac{1}{2}}&{S_0'}
\\{S_c}&{S_0}&{S''}\end{array}\right\}
\left\{ \begin{array}{lll} {T} &{ \frac{1}{2} }&{T_c''}\\
{\frac{1}{2}}&{\frac{1}{2}}&{T_0'}
\\{T_c}&{T_0}&{T''}\end{array}\right\}
\eol
\times 
(S''M_S''S_0'M_{S_0}'| S_c''M_{S_c}'' )
(S_c''M_{S_c}'' \frac{1}{2}\sigma' |SM_S)
(T''M_T''T_0'M_{T_0}'| T_c''M_{T_c}'' )
(T_c''M_{T_c}'' \frac{1}{2}\sigma' |TM_T)
\eol 
\eeqn{wf2}
where $\{\mu\} =\{ M_L''M_S''M_T''\tilde{M}_L\tilde{M}_S'\tilde{M}_T'
\tilde{M}_{S_0}\tilde{M}_{T_0} M_{L_2} m_1\sigma \tau\}$ and
\beq
T^{N\Lambda n_0l_0,N'\Lambda'n_0'l_0'}_{L'',
L_0L_c\kappa l,L_0''L_c''\kappa''l'',L} =
 \sum_{\tilde{N}\tilde{\Lambda}\tilde{n}\tilde{l} n_1l_1 {\cal L} {\cal L}'}
\la \tilde{N}\tilde{\Lambda}\tilde{n}\tilde{l}:{\cal L} | \frac{A(A-3)}{2} | N\Lambda \kappa l:{\cal L} \ra
\la n_1l_1\kappa''l'':{\cal L}' | \frac{A }{A-2} |\tilde{N}\tilde{\Lambda}  n_0l_0:{\cal L}' \ra
\eol \times
\la n_0'l_0'N'\Lambda':L_0''|\frac{A-3}{A-1}|\tilde{n}\tilde{l} n_1l_1:L_0''\ra
\sum_{\lambda_c\lambda_c'}
(\hat{\lambda}_c' \hat{\cal L}' \hat{\lambda}_c \hat{\cal L})^2
\hat{L}_c'' \hat{L}_0'' \hat{L}_c \hat{L}_0
\left\{ \begin{array}{lll} {L}& {l''}&{L_c''}\\{\tilde{l}}&{l_1}&{L_0''}\\
{\lambda_c'}& {{\cal L}'}&{L''}\end{array}\right\}
\left\{ \begin{array}{lll} {L}& {l_0}&{\lambda_c}\\{\tilde{l}}&
{\tilde{\Lambda}}&{{\cal L}}\\
{\lambda_c'}& {{\cal L}'}&{L''}\end{array}\right\}
\left\{ \begin{array}{lll} {L}& {l_0}&{\lambda_c}\\{l}&
{\Lambda}&{{\cal L}}\\
{L_c}& { L_0}&{L''}\end{array}\right\}
\eeqn{}
Then in the
wave function
$\Phi^{(b)}_{K_c'\gamma_c' \kappa' l'LSTM_LM_SM_T}(1,...,A-2,A,A-1)$,
in which the nucleons $A-1$ and $A$ are interchanged,
it is suffucient to extract the pair $(A-2,A)$. This gives
\beq
\Phi^{(b)}_{K_c'\gamma_c' \kappa' l'LSTM_LM_SM_T}(1,...,A-2,A,A-1) =
\sum_{N''\gamma''\tilde{N}'\tilde{\Lambda}' n_0''l_0''L_0''S_0''T_0''
\{\mu'\}}
|A-3 \,N''\gamma''M_L''M_S''M_T''\ra
\eol
\times
\la A-1K_c'\gamma_c'| A-3 N''\gamma'';\tilde{N}'\tilde{\Lambda}',
n_0''l_0''S_0''T_0''( L_0''):
L_c'S_c'T_c'\ra\,
|\psi_{\tilde{N}'\tilde{\Lambda}'}(\ve{\eta}_2')
\psi_{n_0''l_0''}(\ve{\eta}_1'):L_0''M_{L_0}''\ra\varphi_{\kappa'l'm'}(\ve{\xi}_{A-1}')
\eol
\times
\chi_{S_0''M_{S_0}''T_0''M_{T_0}''}(A-2,A)
\chi_{\sigma\tau}(A-1)
(L''M_L''L_0''M_{L_0}''| L_c'M_{L_c}') ( L_c'M_{L_c}' l''m''| LM_L) 
\eol \times
(S''M_S''S_0''M_{S_0}''|S_c'M_{S_c}')
(S_c'M_{S_c}'\frac{1}{2}\sigma |SM_S)
(T''M_T''T_0''M_{T_0}''|T_c'M_{T_c}')
(T_c'M_{T_c}'\frac{1}{2}\tau|TM_T),
\eeqn{wf3}
where $\{\mu'\} = \{ M_L''M_S'' M_T''M_{L_0}''M_{S_0}''M_{T_0}''
M_{L_c}'M_{S_c}'M_{T_c}' m''\sigma\tau\}$. Using Eqs. (\ref{wf2}) and
(\ref{wf3}) in (\ref{ex2}) gives the exchange term $V_{ex}^{(2)}$.
The exchange term $V_{ex}^{(3)}$,
\beq
V_{ex}^{(3)} = 
\la \phi_{\alpha_1'} (1,...,A-2,A) \varphi_{\alpha_2'} (A-1)|
\sum_{i=1}^{A-2}V_{iA-1}|
\phi_{\alpha_1} (1,...,A-2,A-1)\varphi_{\alpha_2} (A)\ra  
\eol
=(A-2)
 \la \phi_{\alpha_1'} (1,...,A) \varphi_{\alpha_2'} (A-1)|V_{A-2,A}|
\phi_{\alpha_1} (1,...,A-2,A-1)\varphi_{\alpha_2}(A )\ra ) 
\eeqn{ex3}
can be calculated in a similar way to provide the final result
\beq 
V_{ex}^{(2)} + V_{ex}^{(3)} =  (A-2)\sum_{ n_0''n_0'l_0' st } 
( W_{n_0''n_0'l_0' st }^{\alpha'\alpha} + W_{n_0'n_0''l_0' st }^{\alpha'\alpha})\,\,
\la\psi_{n_0''l_0'}||V^{st}_{A-2,A}||\psi_{n_0'l_0'}\ra
,
\eeqn{vex23}
where
\beq
W_{n_0''n_0'l_0' st }^{\alpha'\alpha} =
\sum_{N''\gamma'' N\Lambda 
N'\Lambda'n_0l_0 L_0 L_0' S_0S_0'T_0T_0'}
\la A-1K_c\gamma_c| A-3 N''\gamma'';N\Lambda,n_0l_0S_0T_0(L_0):
L_cS_cT_c\ra\,
\eol \times
\la A-1K_c'\gamma_c'| A-3 N''\gamma'';N'\Lambda',
n_0''l_0'st( L_0'):
L_c'S_c'T_c'\ra\,
\eol \times
T^{N\Lambda n_0l_0,N'\Lambda'n_0'l_0'}_{L'',L_0L_c\kappa l,L_0'L_c'\kappa'l',L} \,\,
\hat{S}_c\hat{S}_c'\hat{S}_0\hat{s}
\hat{T}_c\hat{T}_c'\hat{T}_0\hat{t}
\left\{ \begin{array}{lll} {S} &{ \frac{1}{2} }&{S_c'}\\
{\frac{1}{2}}&{\frac{1}{2}}&{s}
\\{S_c}&{S_0}&{S''}\end{array}\right\}
\left\{ \begin{array}{lll} {T} &{ \frac{1}{2} }&{T_c'}\\
{\frac{1}{2}}&{\frac{1}{2}}&{t}
\\{T_c}&{T_0}&{T''}\end{array}\right\}.
\eeqn{}

Finally, the exchange term 
\beq
V_{ex}^{(4)} = 
\la \phi_{\alpha_1'}(1,...,A) 
\varphi_{\alpha_2'}(A-1)|\sum_{i<j}^{A-2}V_{ij}|\phi_{\alpha_1}(1,...,A-2,A-1),
\varphi_{\alpha_2}(A)\ra
\eeqn{vex4}
can be caclulated by separating nucleon $A-1$ in the bra and nucleon $A$
in the ket vector using one-nucleon fractional parentage expansion and
the Talmi-Moshinsky technique. This gives
\beq
V_{ex}^{(4)} = 
\sum_{N''\gamma''\tilde{N}''\tilde{\gamma}''}
\delta_{L''\tilde{L}''}\delta_{S''\tilde{S}''}\delta_{T''\tilde{T}''}
\la A-2 \tilde{N}''\tilde{\gamma}''||\sum_{i<j}^{A-2}V_{ij}|| A-2 N''\gamma''\ra
\,U(S''\frac{1}{2}\frac{1}{2}S;S_cS_c')\,
U(T''\frac{1}{2}\frac{1}{2}T;T_cT_c')
\eol \times
 (-)^{S+S''+S_c+S_c'+T+T''+T_c+T_c'} 
 \sum_{\kappa_1l_1\kappa_1'l_1'} \la A-1 K_c\gamma_c|A-2 N''\gamma'',\kappa_1l_1\ra
 \la A-1 K_c'\gamma_c'|A-2 N''\gamma'',\kappa_1'l_1'\ra
\eol \times 
\sum_{L_0} (-)^{l'+l_1'-L_0}U(L''l_1Ll;L_cL_0)\,U(L''l_1'Ll';L_c'L_0)
 \la \kappa'l'\kappa_1'l_1':L_0|A(A-2)| \kappa_1l_1\kappa l:L_0\ra.
\eol
\eeqn{ex4}
This expression contains the shell model matrix elements of the  
potential energy in the $A-2$ core,
$\la A-2 \tilde{N}''\tilde{\gamma}''||\sum_{i<j}^{A-2}V_{ij}|| 
A-2 N''\gamma''\ra$,
 that can be calculated in the usual way.


\subsection {Overlap integral}

The overlap integral  $I(\ve{r}) = \la\Psi_{A-1}|\Psi_A\ra$ that enters 
reaction theories like the distorted wave Born approximation, coupled reaction
channels,  breakup and capture theories, is a function of distance
between $N$ and the center-of-mass of $A-1$. It is defined in terms of wave
functions $\Psi_{A-1}$ and $\Psi_A$ that depend on non-normalised
Jacobi coordinates $\ve{x}_i = 1/i \sum_{j=1}^i \ve{r}_j - \ve{r}_{i+1}$.
On the contrary, the wave functions in the hyperspherical formalisms 
can be rewritten
in normalised Jacobi coordinates $\ve{\xi}_i = \alpha_i \ve{x}_i$, where
$\alpha_i = \sqrt{i/(i+1)}$. Since
\beq
d\ve{\rho}_A = d\ve{\xi}_1 ...d\ve{\xi}_{A-1} = \prod_{i=1}^{A-1} 
\alpha_i^3 \, d\ve{x}_i,
\eeqn{drho}
the wave function $\Psi_A(\{\ve{x}_i\})$  normalized in coordinates 
$\{\ve{x}_i\}$ is related to the wave function $\tilde{\Psi}_A(\{\ve{\xi}_i\})$
normalized in coordinates $\{\ve{\xi}_i\}$ in the following way:
\beq
\Psi_A(\{\ve{x}_i\}) = \left(\prod_{i=1}^{A-1} \alpha_i^3\right )^{1/2} 
\tilde{\Psi}_A(\{\alpha_i \ve{ x}_i\}) = A^{-3/4} 
\tilde{\Psi}_A(\{\alpha_i \ve{ x}_i\}).
\eeqn{renorm}
Therefore, the overlap integral $I(\ve{x}_{A-1})$ used in reaction theories
is related to the overlap integral $\tilde{I}(\ve{\xi}_{A-1})$
obtained in normalized Jacobi coordinates 
as follows:
\beq
I(\ve{x}_{A-1}) = \int d\ve{x}_1 ...d\ve{x}_{A-2}
\Psi_A^{\dagger}(\{\ve{x}_i\})\Psi_A(\{\ve{x}_i\})
\eol
=
\left(\frac{A-1}{A}\right)^{3/4} \int 
 d\ve{\xi}_1 ...d\ve{\xi}_{A-1}
\tilde{\Psi}_A^{\dagger}(\ve{\xi}_1, ...,\ve{\xi}_{A-2})
\tilde{\Psi}_A(\ve{\xi}_1, ...,\ve{\xi}_{A-2}, \alpha_{A-1}\ve{x}_{A-1}) 
= \left(\frac{A-1}{A}\right)^{3/4} \tilde{I}(\ve{\xi}_{A-1}).
\eeqn{oiren}
 Using the standard for reaction theories partial wave decomposition  
\beq
I_{J_AJ_{A-1}}^{M_AM_{A-1}}(\ve{r})
 =   \sum_{ j m_j l m \sigma }
 (j m_j J_{A-1}M_{A-1}|J_AM_A)
(l m  \frac{1}{2} \sigma  | j m_j)\,
I^{J_AJ_{A-1}}_{lj}(r)\,Y_{lm}(\hat{r})\, 
\chi_{\frac{1}{2}\sigma \frac{1}{2}\tau},
\eeqn{oi}
where $\ve{r} \equiv \ve{\xi}_{A-1}$, $J_i(M_i)$ is the total  momentum 
(its projection) of nucleus $i$, $\chi_{\frac{1}{2}\sigma \frac{1}{2}\tau}$
is the spin-isospin function of the separated nucleon with spin (isospin)
projection $\sigma$ ($\tau$), and expanding the $^5$He
wave function  in a  hyperspherical cluster basis (\ref{y5h3}), we get
 the final expression for the radial part
 of the overlap $\la ^4$He$\otimes n | ^5$He$\ra$,
\beq
I_l(r) = \left(\frac{4}{5}\right)^{5/4}
 r \sum_{n=0}^{n_{\max}} \la K=2n+l | K=0,l\ra 
\int_0^{\infty} d\rho_c \frac{\rho_c^4 \chi_{0}^{^4{\rm He}}(\rho_c) 
\chi_{n}^{^5{\rm He}}\left(\sqrt{\rho_c^2+\frac{4}{5} r^2}\right) 
}{ (\rho_c^2+\frac{4}{5} r^2)^{13/4}}
P_n^{\frac{3}{2},\frac{7}{2}}\left(\frac{\rho_c^2+\frac{4}{5} r^2}
{\rho_c^2+\frac{4}{5} r^2}\right).
\eeqn{oi}
Here $\la K=2n+l | K=0,l\ra $ is the FPC which can be calculated via
the norm of the hyperspherical cluster harmonics. 
The overlap $I_l(r)$ can be represented as a sum of direct and exchange
terms.  The direct term is an analog to the relative
function $g_l(r)$ of the traditional microscopic cluster model 
(\ref{mcm}) and it
can be obtained by removing the antisymmetrization
operator from (\ref{y5h3}). The analytical
expression for this term is obtained from (\ref{oi}) by replacing
the FPC $\la K=2n+l | K=0,l\ra $ by
$5/\la K=2n+l | K=0,l\ra $.

\end{document}